\documentclass
[twocolumn,prb,aps,amsmath,amssymb,floatfix,showpacs]{revtex4-1}

\usepackage{color}
\usepackage{graphicx}
\usepackage{dcolumn}
\usepackage{bm}
\usepackage{subfigure}

\begin{document}

\title{Phonon-limited resistivity of graphene by first-principle calculations: 
electron-phonon interactions, strain-induced gauge field and Boltzmann 
equation}

\author{Thibault Sohier$^{1}$}
\author{Matteo Calandra$^{1}$}
\author{Cheol-Hwan Park$^2$}
\author{Nicola Bonini$^3$}
\author{Nicola Marzari$^4$}
\author{Francesco Mauri$^{1}$}

\affiliation{$^{1}$Institut de Min\'eralogie, de Physique des Mat\'eriaux, 
et de Cosmochimie (IMPMC), Sorbonne Universit\'es - UPMC Univ Paris 06, 
UMR CNRS 7590, Mus\'eum National d'Histoire Naturelle, IRD UMR 206, 
4 Place Jussieu, F-75005 Paris, France.}
\affiliation{$^2$Department of Physics and Center for Theoretical Physics, 
Seoul National University, Seoul 151-747, Korea}
\affiliation{$^3$Department of Physics, King’s College London, 
London WC2R 2LS, UK}
\affiliation{$^4$Theory and Simulation of Materials, 
\'Ecole Polytechnique F\'ed\'erale de Lausanne, 1015 Lausanne, Switzerland}

\date{\today}

\begin{abstract}
We use first-principle calculations, at the density-functional-theory (DFT) 
and GW levels, to study both the electron-phonon interaction for acoustic 
phonons and the ``synthetic'' vector potential induced by a
strain deformation (responsible for an effective magnetic field 
in case of a non-uniform strain).
In particular, the interactions between electrons and 
acoustic phonon modes, the so-called gauge field and deformation potential, 
are calculated at the DFT level in the framework of linear response.
The zero-momentum limit of acoustic phonons is interpreted
as a strain of the crystal unit cell, allowing the calculation of the acoustic 
gauge field parameter (synthetic vector potential) within the GW approximation 
as well.
We find that using an accurate model for the polarizations of the acoustic 
phonon modes is crucial to obtain correct numerical results. 
Similarly, in presence of a strain deformation, the relaxation of atomic 
internal coordinates cannot be neglected.
The role of electronic screening on the electron-phonon matrix elements 
is carefully investigated. 
We then solve the Boltzmann equation semi-analytically in graphene,
including both acoustic and optical phonon scattering.
We show  that, in the Bloch-Gr\"uneisen and equipartition regimes, 
the electronic transport is mainly ruled by the unscreened 
acoustic gauge field, while the  contribution due to the deformation potential 
is negligible and strongly screened. 
We show that the contribution of acoustic phonons to resistivity is doping- and 
substrate-independent, in agreement with experimental observations.
The first-principles calculations, even at the GW level, underestimates this 
contribution to resistivity by $\approx 30 \%$.
At high temperature ($T> 270$ K), the calculated resistivity
underestimates the experimental one more severely, the underestimation being 
larger at lower doping. We show that, beside remote phonon scattering, a
possible explanation for this disagreement is the electron-electron
interaction that strongly renormalizes the coupling to intrinsic optical-phonon
modes. 
Finally, after discussing the validity of the  Matthiessen rule
in graphene, we derive simplified forms of the Boltzmann
equation in the presence of impurities and in a restricted range of
temperatures. These simplified analytical solutions allow us the extract
the coupling to acoustic phonons, related to the strain-induced synthetic 
vector potential, directly from experimental data.
\end{abstract}

\maketitle


\section{Introduction}
\label{sec:Intro}

Electronic transport in graphene has stirred the interest of both
fundamental\cite{Novoselov2005,Zhang2005,DasSarma2011} and applied 
research\cite{Avouris2007} in the past
decade. It provides a unique playground for two-dimensional
carrier dynamics as well as promising technological
breakthroughs. Accurate models and precise understanding of transport
in graphene are thus essential. Intrinsic contributions to resistivity
are of particular interest because they set an ideal limit for
technological improvements to reach. As the fabrication methods
improve, intrinsic contributions begin to dominate the temperature
dependence of  transport measurements\cite{Chen2008,Efetov2010}.
The measured resistivity can now be compared with numerical approaches to the 
intrinsic resistivity.

The dominant contribution to the intrinsic electronic-transport in graphene
comes from the electron-phonon coupling (EPC). Expressions of the EPC matrix 
elements have been derived
\cite{Pietronero1980,Woods2000,Suzuura2002,Piscanec2004,Manes2007,
Samsonidze2007,Venezuela2011} 
and some partial (i.e. including only a restricted set of phonon modes) 
transport models were developed analytically\cite{Shishir2009,Hwang2008a}.
Based on those previous works, the qualitative behavior of acoustic
phonon scattering 
below room-temperature has been successfully determined. 
The low-temperature $\propto T^4$ behavior, typical of 2D electron and
phonon dynamics 
was theoretically predicted\cite{Pietronero1980,Hwang2008a} and experimentally
verified\cite{Efetov2010}, 
as was the linear behavior in the equipartition regime. 
Both those behaviors express the effects of the unique Dirac Cone structure of
graphene. 
Around room-temperature, a remarkable change of behavior in the
temperature dependent resistivity indicates a strong
contribution from a scattering source other than acoustic phonons, often 
attributed to remote optical phonons from the substrate\cite{Chen2008}.

The study of electron-phonon coupling involves the derivation of models 
for the interaction Hamiltonian as well as the phonon spectrum. 
The interaction Hamiltonian was derived within the tight-binding (TB) 
model\cite{Pietronero1980,Woods2000,Suzuura2002,Piscanec2004,
Venezuela2011,Park2014} and in a symmetry-based approach\cite{Manes2007}.
In many works the simple set of strictly longitudinal and transverse phonon 
modes was used. 
However, some qualitative\cite{Manes2007} and 
quantitative\cite{Woods2000,Suzuura2002} models showed that more realistic 
phonon modes 
may be essential to obtain numerically accurate results for acoustic phonon 
scattering. 
EPC parameters have been estimated using ab-initio
simulations\cite{Borysenko2010,Kaasbjerg2012} at the
density-functional theory (DFT) level.
Some combinations of the above models were then inserted in partial transport 
models. 
Overall, the resulting resistivity fell well below experiments, 
due to a lack of completeness and consistency of the EPC and transport models. 
In a previous work\cite{Park2014} we showed that, by calculating the 
resistivity in the framework of the Allen model\cite{Allen1978} 
and including EPC parameters estimated at the GW level\cite{Hedin1965},
a better agreement with experiments could be achieved in the low temperature 
($T < 270$ K), 
high doping regime where acoustic phonon scattering dominates. 
We also noticed a surprisingly important contribution of intrinsic optical 
phonons around room temperature. 
Although their energy is much higher than thermal energy, we found that their 
coupling to electrons 
is much stronger than that of acoustic phonons. This called for further 
investigation 
of this contribution at higher temperatures ($T> 270$ K).

In this work we improve and quantify the most general symmetry-based model 
of EPC via a thorough ab-initio study of the interaction Hamiltonian and the 
phonon modes. 
We link EPC in the long wavelength limit to the perturbation potentials 
induced in strained graphene to enable GW calculations of acoustic EPC 
parameters.
In order to model transport correctly at higher temperatures,
we also go one step beyond in the transport model. We overcome the 
approximations
involved in the Allen model by solving directly the Boltzmann
equation with full inclusion of acoustic and optical phonon modes.
Furthermore we compare our numerical results to experimental data in a larger 
range of temperatures and electron densities.
We show that the resistivity in the equipartition regime is unchanged by 
electron-electron renormalization, and is underestimated by $\approx 30 \%$, at 
all doping levels. 
At high temperatures ($T> 270$ K), the calculated resistivity is dominated by 
intrinsic optical phonons 
and underestimates the experimental one, the underestimation being larger
at lower doping.
Finally we  derive simplified solutions of the Boltzmann
equation in the presence of impurities and valid in a 
restricted range of temperatures.

In Sec. \ref{sec:Ab-initio}, the framework of our ab-initio calculations is 
detailed.
In Sec. \ref{sec:Electrons_Phonons}, we present the Dirac hamiltonian used to
describe the electronic structure and propose a model for phonons modes based 
on ab-initio calculations. 
In Sec. \ref{sec:EPC}, the small-momentum electron-phonon interaction is 
studied analytically and numerically. 
In Sec. \ref{sec:zero-mom}, we develop an interpretation of the zero-momentum 
limit of phonons 
in order to perform GW calculations of EPC parameters.
In Sec. \ref{sec:BTT}, a numerical solution to the
linearized Boltzmann transport equation including all phonon branches is 
developed.
This solution is compared to experiment in Sec. \ref{sec:Results}. 
Finally, in Sec. \ref{sec:App_sol}, semi-analytical approximated 
solutions are presented in order to identify the relevant 
contributions and their relative importance, 
meanwhile proposing more easily implemented numerical solutions. 

\section{Ab-initio Calculations}
\label{sec:Ab-initio}
In this work, we perform density functional theory (DFT) calculations within 
the local lensity approximation\cite{Perdew1981} (LDA)
 using the Quantum-Espresso distribution\cite{Giannozzi2009}.
We use norm-conserving pseudo-potentials with 2s and 2p states in valence and 
cutoff radii of $0.78$ \AA.
We use a $0.01$ Ry Methfessel-Paxton smearing function for the electronic 
integrations and a $65$ Ry kinetic energy cutoff. 
The electron momentum grid depends on the type of calculations performed. 
Accurate band-structures can be obtained at a relatively low computational cost 
with a $16\times 16 \times 1$ electron-momentum grid. 
In the same framework, we used density functional perturbation theory (DFPT) in 
the linear response\cite{Baroni} to perform phonon and electron-phonon coupling 
calculations. In this case, however, a $96\times 96 \times 1$ electron-momentum 
grid was needed to reach convergence.
The distance between graphene and its periodic images is $\approx 20$ \AA.

The GW part of the calculations were done with BerkeleyGW 
package\cite{Deslippe2012}. 
Electronic wave-functions in a $72\times 72 \times 1$ k-point grid are expanded 
in a plane-waves basis with a kinetic energy cutoff of 65 Ry. 
Graphene layers between adjacent supercells are separated by $8.0$ \AA\ and the 
Coulomb interaction is truncated to prevent spurious inter-supercell 
interactions\cite{Ismail-Beigi2006}. The inverse dielectric matrix at zero 
frequency is calculated with a kinetic energy cutoff of 12 Ry and we take into 
account dynamical screening effects in the self energy through the generalized 
plasmon pole model\cite{Hybertsen1986}.

\section{Electrons and Phonons models}
\label{sec:Electrons_Phonons}
\subsection{Dirac Hamiltonian for electrons}
\label{sec:Electrons}

We consider low electron doping of graphene, i.e.  the Fermi level
energy shift from the Dirac point is
$\varepsilon_F \lesssim 0.5$ eV (all energies throughout the paper
are measured with respect to the Dirac point). 
This corresponds to an additional surface charge density of less than 
$1.8 \times 10^{13}$ cm$^{-2}$.
\begin{figure}[h]
\includegraphics[width=0.48\textwidth]{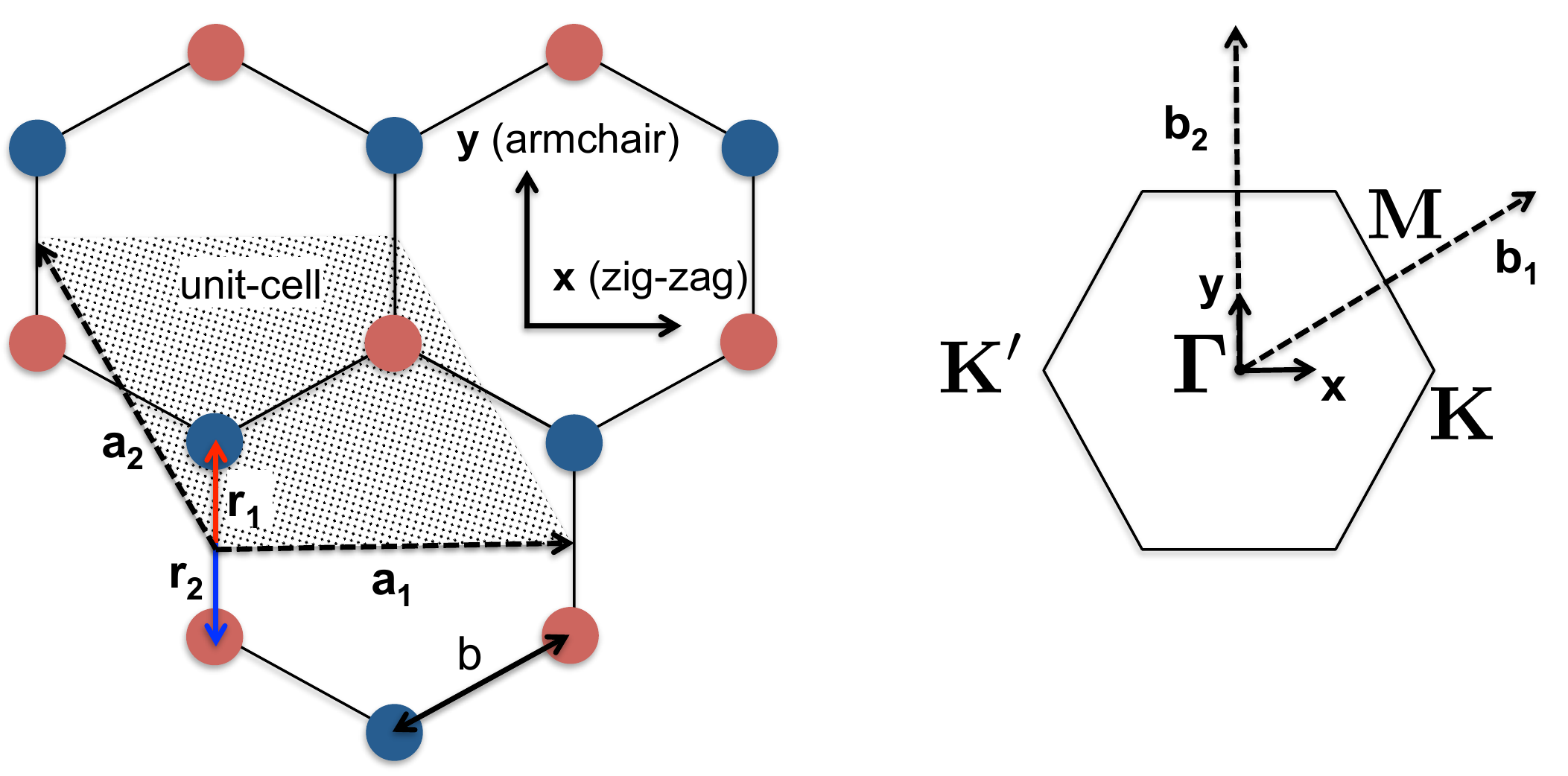}
\caption{(Color online) Definition of the real space unit-cell (left) and the 
first Brillouin zone (right). $\mathbf{a_1}=(1,0)a$ and 
$\mathbf{a_2}=(-1/2,\sqrt{3}/2)a$ are the lattice vectors in Cartesian 
coordinates and $|\mathbf{a_1}|=|\mathbf{a_2}|=a=2.46$ \AA \ is the lattice 
parameter. $b=1.42$ \AA \ is the inter-atomic distance. The $\mathbf{x}$-axis 
($\mathbf{y}$-axis) coressponds to the zig-zag (armchair) direction. 
$\mathbf{r_1}=(1/6,1/3)$ and $\mathbf{r_2}=(-1/6,-1/3)$ are the internal 
coordinates of the carbon atoms in the basis of the lattice vectors. In 
reciprocal space, $\mathbf{b_1}=(1,1/\sqrt{3})2\pi/a$ and 
$\mathbf{b_2}=(0,2/\sqrt{3})2\pi/a$ are the reciprocal lattice vectors in 
cartesian coordinates. In the Brillouin zone, the high symmetry points 
$\mathbf{\Gamma}$, $\mathbf{K}$, $\mathbf{K'}$ and $\mathbf{M}$ are 
represented.}
\label{fig:lattice}
\end{figure}
In this regime, the electronic structure of doped-graphene is well represented 
by two 
Dirac cones\cite{CastroNeto2009} at special points {\bf K}$=(2/3,0)
2\pi/a$ and {\bf K$^{\prime}=(-2/3,0) 2\pi/a$} in Cartesian coordinates, see 
Fig. \ref{fig:lattice}. 
The x-axis is defined as the zig-zag direction of the graphene sheet, and 
$a=2.46$ \AA \ is the
lattice parameter of graphene. We will extend the validity of the Dirac cones 
model to $\varepsilon_F \lesssim 1.0$ eV 
by assuming that the so-called trigonal warping of the bands has a negligible 
effect when the quantities of interest here are angularly averaged.
In the absence of electron-phonon scattering the unperturbed Hamiltonian 
at momentum $\mathbf{k}$ expanded around the Dirac point $\mathbf{K}$ is 
\begin{eqnarray}
\mathcal{H}_{\mathbf{K}}(\mathbf{k})= \hbar v_F
\begin{pmatrix}
 0 &  k_x-ik_y\\
k_x+ik_y  & 0\\
\end{pmatrix}
\label{eq:DiracH}
\end{eqnarray}
where $v_F$ is the Fermi velocity and $\mathbf{k}=(k_x,k_y)$ is the 
electron-momentum  measured with
respect to the Dirac point $\mathbf{K}$, in a Cartesian basis. It can also be 
written as  $\mathcal{H}_{\mathbf{K}}(\mathbf{k})= \hbar v_F \mathbf{k} \cdot 
\mathbf{\sigma}$, where $\mathbf{\sigma}=(\sigma_x, \sigma_y)$ are the Pauli 
matrices.
This Dirac Hamiltonian is written in the pseudospin\cite{DasSarma2011} basis 
emerging from the two inequivalent sub-lattices of graphene. It satisfies the 
eigenvalues equation:
\begin{equation}
\mathcal{H}_{\mathbf{K}}(\mathbf{k})\psi_{\mathbf{k},s}(\mathbf{r})=
\varepsilon_{\mathbf{k},s} \psi_{\mathbf{k},s}(\mathbf{r}) 
\end{equation}
with $\varepsilon_{\mathbf{k},s}=s \ \hbar v_F |\mathbf{k}|$, and $s= \mp 1$ 
for the valence $\pi$ and conduction $\pi^*$ bands respectively.
The Bloch functions are
\begin{equation}
\psi_{\mathbf{k},s}(\mathbf{r})= \frac{1}{\sqrt{N}} e^{i\mathbf{k} \cdot 
\mathbf{r}} | \mathbf{k},s\rangle
\end{equation}
where $N$ is the number of unit-cells in the sample and $| \mathbf{k},s\rangle$ 
is a
pseudospinor eigenfunction, normalized on the unit-cell,
corresponding to the in-plane state $\mathbf{k}$ of the band $s$. 
The eigenfunction $| \mathbf{k},s\rangle$ is defined in the pseudospin basis as:
\begin{equation}
| \mathbf{k},s\rangle= \frac{1}{\sqrt{2S_{\Re}}}  \begin{pmatrix}
e^{-i \theta_{\mathbf{k}} /2} \\
s e^{i \theta_{\mathbf{k} }/2} 
\end{pmatrix}
\label{eq:Dspinor}
\end{equation}
where $S_{\Re}=\frac{\sqrt{3}a^2}{2}$ is the area of a unit-cell. The angle 
$\theta_{\mathbf{k}}$ is the angle between $\mathbf{k}$ and the x-axis.

\subsection{Phonons}

We label $\mathbf{e}_{\mathbf{q},\nu}$ the eigenvector of the dynamical matrix 
corresponding to the phonon mode $\nu$ of momentum $\mathbf{q}$ and eigenvalue 
$\omega_{\mathbf{q},\nu}^2$. This phonon eigenvector is normalized on the 
unit-cell and $\omega_{\mathbf{q},\nu}$ is the frequency of the phonon mode.  
We will discard the coupling to out-of-plane acoustic and optical phonon modes 
since it is zero by symmetry at the linear order\cite{Manes2007}.
The components of the vector $\mathbf{e}_{\mathbf{q},\nu}$ are labeled 
$\mathbf{e}^{a,i}_{\mathbf{q},\nu}$ where $a=1,2$ is an atomic index and 
$i=1,2$ are the in-plane Cartesian coordinates. 
We are particularly interested in the small momentum limit of phonons. 
If we focus on intra-valley scattering, the momentum of phonons that couple to 
electrons is limited by the extension of the Fermi surface, namely 
$|\mathbf{q}| \le 2 k_F$, where $k_F$ is the Fermi wave vector. 
Near the $\mathbf{\Gamma}=(0,0)$ point (i.e. $|\mathbf{q}| \to 0$), it is 
customary to use what will be called here the canonical representation of the 
four in-plane phonon modes to approximate the real ones. The construction of 
those canonical modes relies on the following rules: i) the eigenvector of a 
longitudinal (transverse) mode is parallel (perpendicular) to the phonon's 
momentum; ii) the phase differences between the two atoms of the unit-cell is 
$e^{i \mathbf{q} \cdot (\mathbf{r_1-r_2}) }$ for acoustic modes and 
$-e^{i \mathbf{q} \cdot (\mathbf{r_1-r_2}) }$ for optical modes. This leads to:
\begin{eqnarray}
\mathbf{e}^a_{\mathbf{q},\widetilde{\rm{LA}}} &=& 
\frac{1}{\sqrt{2}}e^{i \mathbf{q} \cdot (\mathbf{R}+\mathbf{r_a}) } 
\frac{\mathbf{q}}{|\mathbf{q}|}  \label{eq:canonical_pola} \\
\mathbf{e}^a_{\mathbf{q},\widetilde{\rm{TA}}} &=& 
\frac{1}{\sqrt{2}} e^{i \mathbf{q} \cdot (\mathbf{R}+\mathbf{r_a})}
\frac{\mathbf{q}_{\perp}}{|\mathbf{q}_{\perp}|} \nonumber \\
\mathbf{e}^a_{\mathbf{q},\widetilde{\rm{LO}}} &=& 
\gamma_a \frac{1}{\sqrt{2}} e^{i \mathbf{q} \cdot (\mathbf{R}+\mathbf{r_a}) } 
\frac{\mathbf{q}}{|\mathbf{q}|} \nonumber \\
\mathbf{e}^a_{\mathbf{q},\widetilde{\rm{TO}}} &=& 
\gamma_a \frac{1}{\sqrt{2}} e^{i \mathbf{q} \cdot (\mathbf{R}+\mathbf{r_a}) } 
\frac{\mathbf{q}_{\perp}}{|\mathbf{q}_{\perp}|} \nonumber
\end{eqnarray}
where $\mathbf{R}$ is the position of the unit-cell, $\mathbf{r_a}$ $a=1,2$ are 
defined in Fig. \ref{fig:lattice} and $\mathbf{q}_{\perp}$ is such that 
$\mathbf{q}_{\perp} \cdot \mathbf{q} =0$. $\gamma_a=\pm 1$ for $a=1,2$ 
respectively. 
The mode indexes $\widetilde{\rm{LA}}$, $\widetilde{\rm{TA}}$ label the 
canonical longitudinal and transverse acoustic phonon modes, respectively. 
The canonical longitudinal and transverse optical phonon modes are labeled 
$\widetilde{\rm{LO}}$ and $\widetilde{\rm{TO}}$ respectively. 

As noted in Ref. \onlinecite{Manes2007}, the real phonon modes of graphene at 
finite momentum tend to the canonical modes in the long wavelength limit. 
However, at finite momentum, there is some mixing between the canonical 
acoustic and optical phonon modes in $o(|\mathbf{q}|)$. We find that the use of 
the canonical eigenvectors leads to a significant error in the following work. 
Therefore we seek an analytical model for the phonon modes that includes 
acoustic-optical mixing. We diagonalize the DFT dynamical matrix, calculated by 
DFT on a small circle around the $\mathbf{\Gamma}$ point. This allows us to 
obtain the angular dependence in $\mathbf{q}$ at fixed $|\mathbf{q}|$. 
Comparing the DFT eigenvectors to the canonical ones, we obtain the following 
expressions:

\begin{eqnarray}
\mathbf{e}_{\mathbf{q},\rm{LA}} &=&   \sqrt{1-\delta^2 |\mathbf{q}|^2} \  
\mathbf{e}_{\mathbf{q},\widetilde{\rm{LA}}} -  \label{eq:DFT_pola}\\
 & &  \delta \   |  \mathbf{q}| \left( \sin(3\theta_{\mathbf{q}}) 
 \mathbf{e}_{\mathbf{q},\widetilde{\rm{LO}}} + \cos(3\theta_{\mathbf{q}}) 
 \mathbf{e}_{\mathbf{q},\widetilde{\rm{TO}}} \right) \nonumber \\
\mathbf{e}_{\mathbf{q},\rm{TA}} &=& \sqrt{1-\delta^2 |\mathbf{q}|^2} \  
\mathbf{e}_{\mathbf{q},\widetilde{\rm{TA}}} +  \nonumber \\ 
 & & \delta \ |\mathbf{q}| \left( \cos(3\theta_{\mathbf{q}}) 
 \mathbf{e}_{\mathbf{q},\widetilde{\rm{LO}}} - \sin(3\theta_{\mathbf{q}}) 
 \mathbf{e}_{\mathbf{q},\widetilde{\rm{TO}}} \right)  \nonumber\\
\mathbf{e}_{\mathbf{q},\rm{LO}} &=&   \sqrt{1-\delta^2 |\mathbf{q}|^2} \  
\mathbf{e}_{\mathbf{q},\widetilde{\rm{LO}}} -  \nonumber \\
 & &  \delta \   |  \mathbf{q}| \left( \sin(3\theta_{\mathbf{q}}) 
 \mathbf{e}_{\mathbf{q},\widetilde{\rm{LA}}} + \cos(3\theta_{\mathbf{q}}) 
 \mathbf{e}_{\mathbf{q},\widetilde{\rm{TA}}} \right)  \nonumber\\
\mathbf{e}_{\mathbf{q},\rm{TO}} &=& \sqrt{1-\delta^2 |\mathbf{q}|^2} \  
\mathbf{e}_{\mathbf{q},\widetilde{\rm{TO}}} +  \nonumber \\ 
  & & \delta  \ |\mathbf{q}| \left( \cos(3\theta_{\mathbf{q}}) 
 \mathbf{e}_{\mathbf{q},\widetilde{\rm{LA}}} - \sin(3\theta_{\mathbf{q}}) 
 \mathbf{e}_{\mathbf{q},\widetilde{\rm{TA}}} \right)  \nonumber
\end{eqnarray}

Where $\delta \approx 0.10$ \AA \ is a small parameter, and 
$\theta_{\mathbf{q}}$ is the angle of $\mathbf{q}$ with respect to the x-axis. 
Our DFT results are consistent with the symmetry-based analysis of Ref. 
\onlinecite{Manes2007}.

In addition to the intra-valley scattering modes at $\Gamma$, we have to 
consider the optical A$_1'$ inter-valley phonon mode, having momentum 
$ \mathbf{K+q}$, with $\mathbf{q}$ being small. The electron-phonon coupling of 
these modes will be parametrized as in Ref. \onlinecite{Piscanec2004}.
\\ 
At small $|\mathbf{q}|$, optical phonons frequencies can be
considered constant ($\hbar \omega_{\rm{TO}}=\hbar \omega_{\rm{LO}}=0.20$ eV, 
$\hbar \omega_{\rm{A}_1'}=0.15$ eV).
Acoustic phonon frequencies are of the form 
$\hbar\omega_{\mathbf{q},\nu}=v_{\nu} |\mathbf{q}|$, where $v_{\nu}$ is the 
sound velocity of $\nu$ mode
(From our DFT calculations, $v_{\rm{TA}}=13.6$ km/s and $v_{\rm{LA}}=21.4$ 
km/s, independent of the direction).
\\

\section{Electron-Phonon matrix elements at finite phonon momentum}
\label{sec:EPC}

In this section we develop the electron-phonon interaction model at small but 
finite momentum (i.e. $|\mathbf{q}| \to 0$), both analytically and numerically.
We use both the canonical and DFT-based eigenvectors of the phonon modes at 
$\mathbf{\Gamma}$, and compare the results.
For the inter-valley scattering A$_1'$ mode at $\mathbf{K}$, the model has 
already been developed\cite{Piscanec2004}, and is simply summarized in 
paragraph \ref{sec:KA_mode}. We will focus on the case of the Hamiltonian 
expanded around the Dirac point $\mathbf{K}$. Similar results are obtained 
around $\mathbf{K'}$ by complex conjugation and the transformations 
$\mathbf{k} \to -\mathbf{k}$ and $\mathbf{q} \to -\mathbf{q}$.
\\

In the basis of Dirac pseudospinors, Eq. \ref{eq:Dspinor}, the small 
$|\mathbf{q}|$ limit of the derivative of the Dirac Hamiltonian with respect to 
a general phonon displacement $\mathbf{e}_{\mathbf{q}}$ gives \cite{Manes2007}:
\begin{equation}
\Delta\mathcal{H}_{\mathbf{q}}= 
\Delta\mathcal{H}_{\mathbf{q},\tilde{A}}
+\Delta\mathcal{H}_{\mathbf{q},\tilde{O}}
\label{eq:cEPC}
\end{equation}
where
\begin{eqnarray}
&&\Delta\mathcal{H}_{\mathbf{q},\tilde{A}} =  i |\mathbf{q}| \times 
\label{Manes_matrixA} \\
 &&\begin{pmatrix}
 2 \alpha(q) Q_{\widetilde{\rm{LA}}} & \tilde{\beta_A} e^{2i\theta_q} 
 (Q_{\widetilde{\rm{LA}}}+iQ_{\widetilde{\rm{TA}}})\\
\tilde{\beta_A} e^{-2i\theta_q} 
(Q_{\widetilde{\rm{LA}}}-iQ_{\widetilde{\rm{TA}}})  & 2\alpha(q) 
Q_{\widetilde{\rm{LA}}}\\
\end{pmatrix} \nonumber
\end{eqnarray}
accounts for the canonical in-plane acoustic modes and
\begin{eqnarray}
&&\Delta\mathcal{H}_{\mathbf{q},\tilde{O}}=i  \times  \label{Manes_matrixO} \\
&&\begin{pmatrix}
 0 & \tilde{\beta_O} e^{i\theta_q} 
 (Q_{\widetilde{\rm{LO}}}+iQ_{\widetilde{\rm{TO}}})\\
\tilde{\beta_O} e^{-i\theta_q} 
(Q_{\widetilde{\rm{LO}}}-iQ_{\widetilde{\rm{TO}}})  &   0 \\
\end{pmatrix} \nonumber
\end{eqnarray}
accounts for the canonical in-plane optical modes. 
Parameters $\tilde{\beta_A}$ and $\tilde{\beta_O}$ are real constants and 
$\alpha(q)$ is a real function of the norm of the phonon momentum 
$q=|\mathbf{q}|$.
The scalar quantities $Q_{\tilde{\nu}}$ are the components of 
$\mathbf{e}_{\mathbf{q}}$ in the basis of the canonical eigenvectors, namely:
\begin{eqnarray}
Q_{\tilde{\nu}}= \mathbf{e}_{\mathbf{q}} \cdot 
\mathbf{e}_{\mathbf{q},\tilde{\nu}}
\end{eqnarray}

$\Delta\mathcal{H}_{\mathbf{q},\nu}$ is easily understood as a change of the 
electronic structure due to the phonon displacement.
In more details:
\begin{itemize}
\item $\beta$-terms (normally labeled ``gauge fields''\cite{Pietronero1980})
in Eqs.  \ref{Manes_matrixA},  \ref{Manes_matrixO} are added to the 
off-diagonal terms
of the Dirac Hamiltonian, Eq. \ref{eq:DiracH}. {\it They shift the Dirac point 
in the Brillouin zone
 without changing its energy}. As such, these terms do not alter
the overall charge and are unaffected by electronic screening.
In a TB model, these terms are related to a variation 
of the nearest neighbors hopping integral with respect to the in-plane lattice 
parameter. 
In  a uni-axially strained graphene sheet, 
the $\beta$-terms correspond to the magnitude 
of the vector potential (the so-called ``synthetic gauge field'' 
\cite{VonOppen2009,Vozmediano2010,Manes2007})
that appear in the perturbed terms of the Dirac Hamiltonian. 
Note that in the presence of a non-uniform strain field, such a synthetic 
vector potential affects the band structure as an effective magnetic 
field\cite{Guinea2010}.

\item $\alpha$-term (labeled ``deformation potential") occurs only in the 
diagonal part of Eq.  \ref{Manes_matrixA}. {\it These terms shift in energy the 
Dirac point, without changing its position in the
Brillouin zone}. As they imply a variation of the charge state, they
are strongly affected by electronic screening. We use here the screened 
deformation potential $\alpha(q)$, in contrast with the original model of Ref. 
\onlinecite{Manes2007} where screening is ignored and a bare constant 
deformation potential $\alpha^{bare}$ is used.
In a TB model this kind of term corresponds to a variation of the on-site 
energy. In mechanically strained graphene, 
it represents the magnitude of the scalar potential or 
``synthetic electric field'' \cite{VonOppen2009,Manes2007} triggered by a 
change in the unit cell area. 
\end{itemize}

The EPC matrix elements  are defined as 
\begin{eqnarray}
g_{\mathbf{k}+\mathbf{q},s,\mathbf{k},s',\nu}
 &=& \sqrt{\frac{\hbar}{2M\omega_{\mathbf{q},\nu}}}  \langle  \mathbf{k+q},s |
 \Delta\mathcal{H}_{\mathbf{q},\nu}|\mathbf{k},s' \rangle  
\label{eq:defg}
\end{eqnarray}
where $M$ is the mass of a carbon atom, $| \mathbf{k},s' \rangle $ and 
$|\mathbf{k+q},s \rangle$ are the initial and final electronic states of the 
scattering process. 
Since most scattering processes significantly contributing to transport are 
intra-band, we will drop the $s$ and set $s=s'=1$ unless specified otherwise. 
Setting $s=s'=-1$ would give the same final results due to electron-hole 
symmetry. We further simplify the notation by setting :
$$\langle  \mathbf{k+q}, 1|\Delta\mathcal{H}_{\mathbf{q},\nu}
|\mathbf{k}, 1 \rangle= \Delta\mathcal{H}_{\mathbf{q},\nu}(\mathbf{k})$$
We will now study the EPC models obtained using either canonical or DFT phonon 
modes at $\mathbf{\Gamma}$ with Eq. \ref{eq:cEPC}.

\subsection{Coupling to canonical phonon modes at $\mathbf{\Gamma}$}

Using the canonical phonon modes, the small phonon-momentum limit 
($|\mathbf{q}|\to 0$) of $\Delta\mathcal{H}_{\mathbf{q},\nu}(\mathbf{k})$ can 
be written as:
\begin{eqnarray}
 |\Delta\mathcal{H}_{\mathbf{q},\widetilde{\rm{TA}}}(\mathbf{k})|&=& 
 \left| \tilde{\beta_A} |\mathbf{q}| \sin\left(2\theta_{\mathbf{q}}
 +\frac{\theta_{\mathbf{k+q}}+\theta_{\mathbf{k}} }{2}\right) \right|
\label{eq:DcTA}\\
 |\Delta\mathcal{H}_{\mathbf{q},\widetilde{\rm{LA}}}(\mathbf{k})|&=& \bigg| 2 
 \alpha(q)  |\mathbf{q}| \cos\left( 
 \frac{\theta_{\mathbf{k+q}}-\theta_{\mathbf{k}} }{2} \right) \nonumber \\
&+& \tilde{\beta_A} |\mathbf{q}| \cos \left( 
2\theta_{\mathbf{q}}+
\frac{\theta_{\mathbf{k+q}}+\theta_{\mathbf{k}} }{2}\right) \bigg| 
\label{eq:DcLA}\\
 |\Delta\mathcal{H}_{\mathbf{q},\widetilde{\rm{LO}}}(\mathbf{k})|&=& 
 \left|\tilde{\beta_O}  \sin \left(\theta_{\mathbf{q}} - 
 \frac{\theta_{\mathbf{k+q}}+\theta_{\mathbf{k}} }{2} \right) \right| 
\label{eq:DcLO}\\
 |\Delta\mathcal{H}_{\mathbf{q},\widetilde{\rm{TO}}}(\mathbf{k})| &=& 
 \left|\tilde{\beta_O}  \cos \left(\theta_{\mathbf{q}}- 
 \frac{\theta_{\mathbf{k+q}}+\theta_{\mathbf{k}} }{2}\right) \right|  
\label{eq:DcTO}
\end{eqnarray}
These expressions were obtained by symmetry considerations  in Ref. 
\onlinecite{Manes2007}. Using a TB model  \cite{Venezuela2011,Piscanec2004,
Pietronero1980,Park2014}, similar expressions can be obtained.
Due to their high energy, $\widetilde{\rm{LO}}$ and $\widetilde{\rm{TO}}$ 
phonon modes can involve inter-band ($\pi-\pi^*$) scattering. However, setting 
$s=1$ and $s'=-1$ simply exchanges the angular dependencies of 
$\widetilde{\rm{TO}}$ and $\widetilde{\rm{LO}}$. This has no impact in the 
following transport model since the contributions of those modes are always 
summed. We can thus keep the above expressions without loss of generality.

\subsection{Coupling to DFT phonon modes at $\mathbf{\Gamma}$}

We now insert the DFT phonon modes, Eq. \ref{eq:DFT_pola}, into the canonical 
model of Eq. \ref{eq:cEPC}. 
In the expressions for the acoustic DFT eigenvectors (Eq. \ref{eq:DFT_pola}), 
the angular dependency and $|\mathbf{q}|$ behavior of the $\widetilde{\rm{TO}}$ 
and $\widetilde{\rm{LO}}$ components are of the same form as in Eqs. 
\ref{eq:DcTA} and \ref{eq:DcLA}, when considering a circular Fermi surface.
It can be shown that the effect of the DFT phonons eigenvectors model on 
$\Delta\mathcal{H}_{\mathbf{q},\rm{TA}}(\mathbf{k})$ and 
$\Delta\mathcal{H}_{\mathbf{q},\rm{LA}}(\mathbf{k})$ is then a simple 
redefinition of the magnitude $\tilde{\beta_A}$. 
Concerning the matrix elements derived from the DFT optical modes, the 
contribution of the canonical acoustic modes is in $o(|\mathbf{q}|^2)$ and can 
be neglected with respect to the dominant $o(1)$ term from the canonical 
optical modes. Optical EPC matrix elements are thus unaffected by the ab-initio 
model for phonons. 

The small phonon-momentum limit ($|\mathbf{q}| \to 0$) of 
$\Delta\mathcal{H}_{\mathbf{q},\nu}(\mathbf{k})$ for the DFT modes can then be 
written :

\begin{eqnarray}
|\Delta\mathcal{H}_{\mathbf{q},\rm{TA}}(\mathbf{k})|&=& \left| \beta_A 
|\mathbf{q}|  \sin \left( 
2\theta_{\mathbf{q}}+\frac{\theta_{\mathbf{k+q}}
+\theta_{\mathbf{k}} }{2}\right) \right|
\label{eq:DTA}\\
|\Delta\mathcal{H}_{\mathbf{q},\rm{LA}}(\mathbf{k})|&=& \bigg| 2 \alpha(q)  
|\mathbf{q}| \cos\left( \frac{ \theta_{\mathbf{k+q}}-\theta_{\mathbf{k}} }{2} 
\right) \nonumber \\
  &+ & \beta_A |\mathbf{q}| \cos \left( 
2\theta_{\mathbf{q}}+\frac{\theta_{\mathbf{k+q}}+
\theta_{\mathbf{k}} }{2}\right) \bigg| \label{eq:DLA}\\
|\Delta\mathcal{H}_{\mathbf{q},\rm{LO}}(\mathbf{k})|&=& \left|\beta_O  \sin 
\left(\frac{\theta_{\mathbf{k+q}}+
\theta_{\mathbf{k}} }{2}- \theta_{\mathbf{q}}\right) \right| 
\label{eq:DLO}\\
|\Delta\mathcal{H}_{\mathbf{q},\rm{TO}}(\mathbf{k})|&=& \left|\beta_O  
\cos \left(\frac{\theta_{\mathbf{k+q}}+
\theta_{\mathbf{k}} }{2}- \theta_{\mathbf{q}}\right) \right|  
\label{eq:DTO}
\end{eqnarray}

Where $\beta_A= \sqrt{1-\delta^2|\mathbf{q}|^2}\tilde{\beta_A}- \delta 
\tilde{\beta_O} \approx \tilde{\beta_A}- \delta \tilde{\beta_O}$ and 
$\beta_O\approx \tilde{\beta_O}$ are effective parameters, and $\alpha(q)$ is 
unchanged because there are no diagonal terms in Eq. \ref{Manes_matrixO}. 

\subsection{Coupling to inter-valley A$_1'$ mode at $\mathbf{K}$}
\label{sec:KA_mode}

The inter-valley A$_1'$ mode at $\mathbf{K}$ scatters an electron from  state 
$\mathbf{k}$ around $\mathbf{K}$ to state $\mathbf{k+q}$ around $\mathbf{K'}$, 
where $\mathbf{q}$ is still small.
Using a TB model, it is found to be\cite{Piscanec2004}:
\begin{equation}
|\Delta\mathcal{H}_{\mathbf{K+q},\rm{A}_1'}(\mathbf{k})| =  \left| \sqrt{2} 
\beta_{K}  \sin\left( 
\frac{ \theta_{\mathbf{k+q}}-\theta_{\mathbf{k}} }{2} \right) \right| 
\label{eq:DKA1}
\end{equation}
where $\beta_K$ is a real constant.
This high energy mode will involve inter-band scattering. In the case $s=-s'$, 
the above expression becomes:
\begin{equation}
|\langle  \mathbf{k+q}|
\Delta\mathcal{H}_{\mathbf{K+q},\rm{A}_1'}^{s=-s'}|\mathbf{k} \rangle| =  
\left| \sqrt{2} \beta_{K}  \cos\left( 
\frac{ \theta_{\mathbf{k+q}}-\theta_{\mathbf{k}} }{2} \right) \right| 
\label{eq:DKA1}
\end{equation}

\subsection{Calculation of EPC parameters from DFPT in the linear response.}
\label{sec:DFPT_num_EPC}

In this section we perform direct ab-initio calculations of acoustic EPC matrix 
elements by using density functional perturbation theory\cite{Baroni}.
The parameters $\beta_O, \beta_K$ for optical phonons have already been 
evaluated using this method\cite{Piscanec2004,Venezuela2011} and compared to 
experimental Raman measurements. 
Their numerical values are reported in Table \ref{EPC_param}. 
We will mainly focus here on the acoustic phonon EPC parameters.

It is important to underline that this technique do not provide access to the 
bare deformation potential parameter $\alpha^{bare}$ due to electronic 
screening. 
In linear response at small finite (i.e. non-zero) phonon
momentum $\mathbf{q}$, the phonon displacement induces a small but finite 
$\mathbf{q}$-modulated electric field. 
The electrons screen the finite electric field and
consequently the magnitude of EPC. 
Thus, a finite induced electric field is always present in the linear
response calculation at any non-zero phonon momentum
and the screened parameter $\alpha(q)$ is obtained. 

Concerning the gauge field terms, both the canonical ($\tilde{\beta_A}$) and 
effective ($\beta_A$) EPC parameters have been calculated to verify the 
consistency of our model.

\begin{figure}[h]
\includegraphics{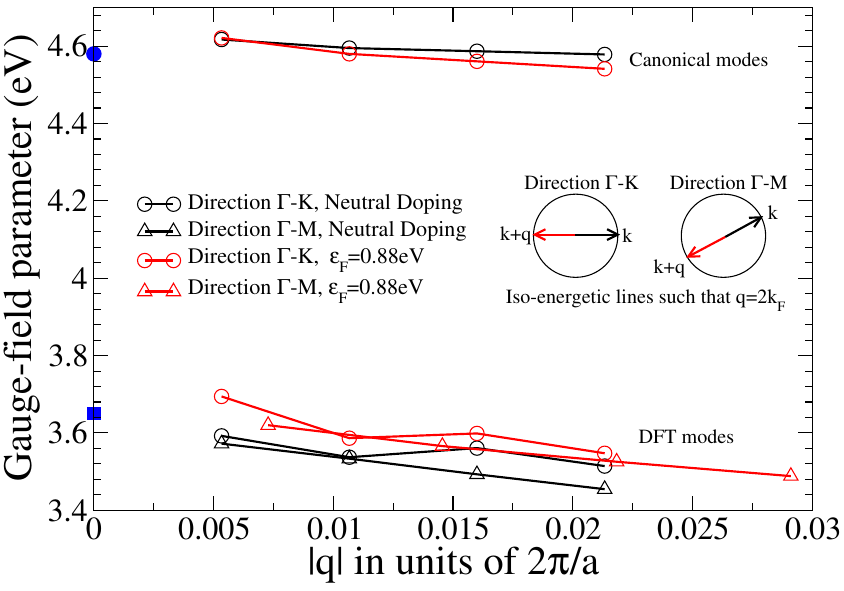} 
\caption{(Color online)Ab-initio calculations of 
$\frac{|\Delta\mathcal{H}_{\mathbf{q},\rm{LA/TA}}(\mathbf{k})|}{|\mathbf{q}|}$ 
in high symmetry 
$(\mathbf{\Gamma} \to \mathbf{M}) / (\mathbf{\Gamma} \to \mathbf{K})$ 
directions respectively lead to gauge field EPC parameters, which depend 
neither on direction nor doping. $\tilde{\beta_A}$($\beta_A$) is obtained when 
the canonical (DFT) phonon modes are used. The blue square and dot at 
$|\mathbf{q}|=0$ represent the values of $\beta_A$ and $\tilde{\beta_A}$ found 
in strained graphene calculations, respectively.}
\label{fig:beta}
\end{figure}

By choosing the phonon momentum ${\bf q}$ along the high symmetry directions 
$\mathbf{\Gamma} \to \mathbf{K}$ and $\mathbf{\Gamma} \to \mathbf{M}$, we have 
 $\theta_q= 0$ and $\pi/6$, respectively. If initial and scattered states are 
 taken on a circular iso-energetic line, i.e. if $|\mathbf{k}|=|\mathbf{k+q}|$, 
 then $\frac{\theta_{\mathbf{k+q}}+\theta_{\mathbf{k}} }{2}=\theta_{\mathbf{q}} 
 \pm \frac{\pi}{2}$. From Eqs.  \ref{eq:DTA} and \ref{eq:DLA}
we obtain:

\begin{eqnarray}
|\Delta\mathcal{H}^{(\mathbf{\Gamma \to K})}_{\mathbf{q},\rm{TA}}(\mathbf{k})|
&=& 
|\mathbf{q}| \beta_A \label{eq:DGKTA}\\ 
|\Delta\mathcal{H}^{(\mathbf{\Gamma \to K})}_{\mathbf{q},\rm{LA}}(\mathbf{k})| 
&=& |\mathbf{q}| \left|2 \alpha(q) \cos \left(
\frac{\theta_{\mathbf{k+q}} -\theta_{\mathbf{q}} }{2}  \right) \right| 
\label{eq:DGKLA}\\
|\Delta\mathcal{H}^{(\mathbf{\Gamma \to M})}_{\mathbf{q},\rm{TA}}(\mathbf{k})|
&=& 0 \label{eq:DGMTA}\\
|\Delta\mathcal{H}^{(\mathbf{\Gamma \to M})}_{\mathbf{q},\rm{LA}}(\mathbf{k})| 
&=& |\mathbf{q}| \left| \pm \beta_A+2 \alpha(q)  \cos \left(    
\frac{\theta_{\mathbf{k+q}} -\theta_{\mathbf{q}} }{2}  \right) \right| 
\nonumber\\  \label{eq:DGMLA}
\end{eqnarray}
We then consider an iso-energetic line at $\varepsilon=\hbar v_F 
|\mathbf{q}|/2$ 
and select the electron-momentum ${\bf k}$-point such that 
$\theta_{\mathbf{k+q}} -\theta_{\mathbf{q}}=\pi$. 
In this way the cosines in Eqs. \ref{eq:DGKLA} and \ref{eq:DGMLA}  are null 
and only the contribution of $\beta_A$ coefficient remains. 
Although we used the notations of the effective model here, the same strategy 
can be applied to the canonical model. 
The EPC matrix element is then calculated in either the canonical or DFT 
eigenvectors basis to obtain $\tilde{\beta_A}$ or $\beta_A$, respectively.
In Fig. \ref{fig:beta}, we plot the resulting $\tilde{\beta_A},\beta_A$ for 
different doping conditions. 
The fact that the results are the same if evaluated for ${\bf q}$ along 
$\mathbf{\Gamma} \to \mathbf{K}$ or $\mathbf{\Gamma} \to \mathbf{M}$
confirms the angular dependencies of Eqs.  \ref{eq:DTA} and \ref{eq:DLA}. As 
expected, the gauge field terms
 $\tilde{\beta_A}$ and $\beta_A$ are essentially doping independent and 
 screening has no effect on it. 
A direct consequence is that scattering by gauge-field is independent of the 
dielectric background and thus independent of the substrate.
The numerical results are reported in the first column of Table \ref{EPC_param}.

Knowing the value of $\tilde{\beta_A}$ and $\beta_A$, we adopt a similar 
strategy to obtain the {\it screened} $\alpha(q)$ coefficient by setting 
$\theta_{\mathbf{k+q}} -\theta_{\mathbf{q}}=2\pi/3$.
We find that the $\alpha(q)$ coefficient is smaller than the numeric noise of 
our simulation. 
It is then completely negligible with respect to the other parameters. 

\begin{table*}
\caption{ Electron-Phonon Coupling parameters, calculated by different methods: 
i) DFPT EPC: direct DFPT LDA calculation of EPC, Sec. \ref{sec:DFPT_num_EPC}. 
This method does not give access to unscreened $\alpha^{bare}$. 
ii) $|\mathbf{q}|=0$: from zero-momentum model, see Sec. \ref{sec:zero-mom}. 
Acoustic parameters are obtained by calculating the magnitude of 
strained-induced scalar and vector potentials. Optical parameters are obtained 
with the frozen phonon method from Ref. \onlinecite{Lazzeri2008}.
iii)TB-DFT: Results obtained in our previous work\cite{Park2014} using a TB 
model and DFT to calculate the derivative of the hopping parameter with respect 
to bond-length (see Sec. \ref{sec:TB-EPC}). The deformation potential 
$\alpha^{bare}$ was set to zero in this work.
iv) GW calculations of EPC parameters. For phonons at $\mathbf{\Gamma}$, the 
renormalization is 20$\%$, as the Fermi velocity. For the A$_1'$ mode, results 
are taken from Ref. \onlinecite{Attaccalite2010}.
v) Exp: Obtained by fitting our numerical solution of Boltzmann transport 
equation to experiment. $\tilde{\beta_A}$ and $\alpha^{bare}$ are not used in 
the simulations. The values of $\beta_K$ in the last line are doping dependent, 
see Sec. \ref{sec:Results} for plots (Fig. \ref{fig:betaK}) and discussion.\\ 
Note that in a nearest neighbor TB model (third column) $\beta_K=\beta_O$.
This is verified\cite{Piscanec2004} within a $1\%$ error in the frameworks 
of the first two columns. At the GW level however, this equality does not hold.}
\begin{center}
\begin{tabular*}{0.6\textwidth}{ l | c | c | c | c | c   }
\hline 
\hline
&DFPT EPC  & $|\mathbf{q}|=0$ &  TB-DFT\cite{Park2014} & GW & Exp    \\
\hline 
$\tilde{\beta_A}$ (eV) & 4.60  &  4.58 & 3.54 &  5.52  & --      \\
$\beta_A$ (eV)  & 3.60 &  3.64  & 3.54   &  4.32    & 4.97              \\
$\alpha^{bare}$(eV)    & --      &    2.96     &   -- & --   & --         \\
$\beta_O=\tilde{\beta_O}$ (eV/\AA) & 9.5 &  9.5 &  10.08 & 11.39 & 11.39   \\
$\beta_K$ (eV/\AA)  & 9.5 & 9.5 &  10.08 &  12.5 $ \sim$ 14  & 17 $\sim$ 40  \\
\end{tabular*}
\end{center}
\label{EPC_param}
\end{table*} 

\subsection{EPC parameters in the tight-binding model}
\label{sec:TB-EPC}
In this section we compare the results of DFPT with other results obtained 
within the TB model.

In our previous work\cite{Park2014}, the canonical phonon modes were used to 
calculate the perturbation to the TB Hamiltonian. Thus, the canonical EPC 
parameters $\widetilde{\beta_A}$, $\widetilde{\beta_O}$ were obtained. In the 
TB model, $\widetilde{\beta_A}$, $\widetilde{\beta_O}$ and $\beta_K$ are all 
proportional to the derivative $\eta$ of the nearest neighbor hopping integral 
with respect to bond-length. Such relationships are obviously very specific to 
the TB model and are not enforced in the symmetry-based model used here. DFT 
calculations of $\eta$ resulted in the numerical values of EPC parameters 
reported in Table \ref{EPC_param}, column "TB-DFT". Those values were also 
checked against DFPT calculations similar to those presented here. The acoustic 
EPC obtained in DFPT calculations, using the DFT phonon modes, resulted in what 
we call here $\beta_A$, although the distinction was not made at that time. 
Indeed, the value of $\widetilde{\beta_A}$ in the TB model happens to be close 
to the values of $\beta_A$ found by DFPT. This leads to a very good numerical 
agreement between the low temperature resistivity calculated in this work and 
the previous one. However, we would like to point out that this agreement is 
fortuitous, and in view of the analysis made here, it should be interpreted as 
a manifestation of the limits of the TB model.

An other way to evaluate EPC matrix elements for acoustic phonons can be found 
in Refs. \onlinecite{Woods2000,Suzuura2002}. The Hamiltonian for 
electron-phonon is similarly derived from a TB model. The important difference 
is the use of a microscopic "valence-force-field"(VFF) model to derive the 
dynamical matrix and the resulting phonon modes. This model involves two 
parameters describing the forces resulting from changes in bond-lengths and 
angles in the lattice. Those parameters are fitted on graphene's phonon 
dispersion derived from models using the force-constants measured in graphite. 
The resulting phonon modes are then inserted in the TB electron-phonon 
interaction Hamiltonian. Results qualitatively similar to our work are 
obtained. In particular, a reduction factor originating from the mixing of the 
acoustic and optical canonical modes appears in the coupling to acoustic modes. 
The parameter obtained in our DFPT calculations using the canonical phonon 
modes ($\widetilde{\beta_A}$) falls in the interval estimated in Refs. 
\onlinecite{Woods2000,Suzuura2002}. However, the aforementioned reduction 
factor, equivalent to the ratio $\beta_A/\widetilde{\beta_A}$, is found to be 
$\approx 0.5$ in Refs. \onlinecite{Woods2000,Suzuura2002} while we find 
$\beta_A/\widetilde{\beta_A} \approx 0.78$.

\section{Electron-phonon coupling at zone center from finite deformations}
\label{sec:zero-mom}

In order to calculate the electron-phonon matrix elements in the GW 
approximation, we calculate the GW band structure for suitably chosen 
deformation patterns.
If the displacement pattern is chosen to reproduce the zero-momentum limit of a 
given phonon, the matrix elements of the resulting perturbation Hamiltonian can 
be linked to the EPC parameters. 
Following this approach, the frozen phonons method\cite{Lazzeri2008} was used 
to calculate 
the electron-electron renormalization of the coupling to {\it optical } modes 
(LO, TO, A$_1'$) within GW.
In order to perform GW calculations and to check the consistency of the small 
momentum EPC model, 
we also seek an interpretation of the {\it acoustic} EPC parameters at momentum 
exactly zero. 
This is achieved by linking acoustic EPC parameters to the perturbation 
potentials induced by a mechanical strain. 
This link is then verified numerically at the DFT level. Finally, we present 
the results of GW calculations for acoustic EPC parameters using this method, 
and summarize the already existing results on the optical EPC parameters.

\subsection{Acoustic EPC and strain-induced potentials}
\label{EPC-strain}

For acoustic phonons at $\Gamma$, a static phonon displacement in the zero 
momentum limit is equivalent to a strain deformation. 
We consider the long wavelength limit of an acoustic phonon and the 
corresponding perturbation occurring on a portion 
of the graphene sheet of scale $d << \frac{2 \pi}{q}$. Provided there is no 
long-range (Coulomb) interaction between such zones distant to each other,
the phonon perturbation can be seen locally as a simple mechanical strain of 
the sheet. 
Author of Ref. \onlinecite{Manes2007} derived the $|\mathbf{q}| \to 0$ limit of 
the electron-phonon interaction (Eqs. \ref{eq:DcTA} and \ref{eq:DcLA}) for the 
canonical acoustic modes presented in Sec. \ref{sec:EPC} (Eq. 
\ref{eq:canonical_pola}). 
Since screening was ignored in this EPC model, the magnitude of the {\it bare} 
deformation potential $\alpha^{bare}$ was used.
In this framework, there is no long-range interaction and strain can be 
considered 
to be the exactly $|\mathbf{q}|=0$ equivalent of the $|\mathbf{q}| \to 0$ limit 
of an acoustic phonon. 
We will discuss the consequences of screening on the interpretation of the 
deformation potential in the zero-momentum limit in paragraph \ref{sec:alpha}.
We first review the model of strain introduced in Ref. \onlinecite{Manes2007}. 
This model will be called canonical.
The strained unit-cell is defined with the lattice vectors 
$\mathbf{a'_1}, \mathbf{a'_2}$ such that :
\begin{eqnarray}
\mathbf{a'_i}&=&(\mathcal{I}+\mathcal{U})\  \mathbf{a_i} \\
\mathcal{U}&=& \begin{pmatrix}  u_{xx} & u_{xy}  \\  u_{yx} & u_{yy} 
\end{pmatrix}
\end{eqnarray}
where $\mathcal{I}$ and $\mathcal{U}$ are the identity matrix and strain 
tensor, respectively.
In this first canonical model, the vectors defining the positions of the carbon 
atoms in real space are given by the same transformation as the lattice vectors.
Namely, the internal coordinates of atoms are unchanged in the basis of the 
lattice vectors $\{\mathbf{a'_1},\mathbf{a'_2}\}$.
Evidently, strain also changes the reciprocal lattice vectors according to the 
transformation $\mathbf{b'_i}= (\mathcal{I}+\mathcal{U})^{-1} \mathbf{b_i}$. 
It is then natural to develop the Hamiltonian around the special point 
$\mathbf{K}=(2/3,-1/3)$, as defined in the basis of those new reciprocal 
lattice vectors. While the coordinates of $\mathbf{K}$ in the basis of the 
reciprocal lattice vectors are unchanged, it is useful for DFT calculations to 
note that the Cartesian coordinates of this special point changed compared to 
the unstrained case. This change is only due to the geometrical redefinition of 
the lattice vectors. 
The canonical perturbed Hamiltonian 
$\mathcal{\widetilde{H}}^{\mathcal{U}}_\mathbf{K}$, expanded around point 
$\mathbf{K}$ of strained graphene is then written in terms of a vector 
potential $\mathbf{\tilde{A}}=(\tilde{A}_x,\tilde{A}_y)$ and a scalar potential 
$\Phi$\cite{Manes2007}:
\begin{eqnarray}
\mathcal{\widetilde{H}}^{\mathcal{U}}_\mathbf{K} (\mathbf{k})= \Phi \ 
\mathcal{I} + (\hbar v_F \mathbf{k}+\mathbf{\tilde{A}}) \cdot \mathbf{\sigma}  
\label{eq:Hstrain-beg} 
\end{eqnarray}
where
\begin{eqnarray}
\tilde{A}_x &=& \sqrt{2} \tilde{\beta_A} (u_{xx}-u_{yy}) \\
\tilde{A}_y &=& -\sqrt{2} \tilde{\beta_A} (u_{xy}+u_{yx}) \\
\Phi &=& \sqrt{2} \alpha^{bare} (u_{xx}+u_{yy}) 
\label{eq:Hstrain-end}
\end{eqnarray} 
$\Phi$ acts as a global energy shift while $\mathbf{\tilde{A}}$ yields a 
redefinition of the Dirac point's position. In strained graphene and in the 
presence of gauge fields, it is then important to distinguish special point 
$\mathbf{K}$ from the Dirac point labeled $\mathbf{\bar{K}}$ here. The former 
is defined geometrically while the latter is defined as where the $\pi$ and 
$\pi^*$ bands intersect. 
The Dirac point is now $\mathbf{\bar{K}}= \mathbf{K} - 
\frac{\mathbf{\tilde{A}}}{\hbar v_F}$, as can be seen in Eq. 
\ref{eq:Hstrain-beg}. 
Similar expressions are obtained around the other Dirac cone $\mathbf{K'}$, by 
complex conjugation and the transformations $\mathbf{k} \to -\mathbf{k}$ and 
$\mathbf{\tilde{A}} \to -\mathbf{\tilde{A}} $.

An addition to the canonical model of EPC Eqs. \ref{eq:DcTA} and \ref{eq:DcLA} 
was the use of the DFT phonon modes of Eq. \ref{eq:DFT_pola}. In order to 
obtain the strain pattern equivalent to the $|\mathbf{q}| = 0$ limit of those 
modes, we allow the relaxation of internal coordinates after imposing a given 
strain to the crystal axes. This structural optimization of the internal 
coordinates is crucial as after the strain deformation, there are non-zero 
forces on the atoms.
To give some numerical example, consider the positions of the two carbon atoms 
as defined in Fig. \ref{fig:lattice}. If we apply to the unit-cell a $1\%$ 
uni-axial strain in the $\mathbf{y}$-direction,
\begin{equation}
\mathcal{U}= \begin{pmatrix}  0 & 0 \\ 0  & 0.01 \end{pmatrix}
\end{equation}
and allow the relaxation of internal coordinates, we obtain the new atomic 
positions $\mathbf{r'_1}= -\mathbf{r'_2} \approx 0.9987 \times (1/6, 1/3) $ in 
the $\{ \mathbf{a'_1},\mathbf{a'_2} \}$ basis.
As was the case for phonon modes at small momentum, this relaxation has 
substantial numerical consequences. 
We assume that the internal coordinates relaxation process leads to a strain 
model with gauge field parameter $\beta_A$ (as in Eqs. \ref{eq:DTA} and 
\ref{eq:DLA}) instead of $\tilde{\beta_A}$ (as in Eqs. \ref{eq:DcTA} and 
\ref{eq:DcLA}). This is analogous to the effects of using the DFT phonon modes 
of Eq. \ref{eq:DFT_pola} in DFPT.
We will thus use the following effective strain model :
\begin{eqnarray}
\mathcal{H}^{\mathcal{U}}_\mathbf{K} (\mathbf{k}) &=& \Phi \ \mathcal{I} + 
(\hbar v_F \mathbf{k}+\mathbf{A}) \cdot \mathbf{\sigma} \\
A_x &=& \sqrt{2} \beta_A (u_{xx}-u_{yy}) \\
A_y &=& -\sqrt{2} \beta_A (u_{xy}+u_{yx}) \\
\Phi &=& \sqrt{2} \alpha^{bare} (u_{xx}+u_{yy}) 
\label{eq:Hstrain2-end}
\end{eqnarray}  

\subsection{Calculation of strain-induced potentials at the DFT level}
\label{sec:strain-Num}

In this section we calculate the changes in the electronic structure of 
graphene\cite{Choi2010} under strain within DFT.
The magnitudes of the scalar and vector potentials are then extracted from the 
displacement in the Brillouin
zone and the energy shift of the Dirac cone. Within DFT, doping has a 
negligible effect on this process. By comparing the results to the previous 
DFPT results, we validate the zero-momentum strain model.

\subsubsection{Calculation and interpretation of the strain-induced bare 
deformation potential}
\label{sec:alpha}

If there are no long-range interactions then the global energy shift $\Phi$ 
plays the role of the bare deformation potential part of the EPC at 
$|\mathbf{q}| = 0$.
Since long-range Coulomb interactions and screening are present in our EPC 
model, however, an additional complication appears.
In finite difference calculations (DFT calculations on strained graphene), 
differently to what happen in DFPT, the diagonal perturbation $\Phi$ shifts the 
Dirac Point in energy just by adding a constant potential with no modulation in 
$\mathbf{q}$.
As a result, the scalar potential $\Phi$ obtained with finite differences is 
bare and does not correspond to the $|\mathbf{q}| \to 0 $ limit of the screened 
deformation potential used in our EPC model.
We can obtain the screened $\alpha(q)$ by assuming :
\begin{equation}
\alpha(q) = \frac{\alpha^{bare} }{\epsilon(q)}
\end{equation}
where $\epsilon(q)$ is the static dielectric function of graphene in the 
random-phase approximation\cite{Ando2006,Hwang2008}.

The parameter $\alpha^{bare}$ is obtained by a global variation of bond-length 
$b$ and is related to the scalar potential $\Phi$:
\begin{eqnarray}
\mathcal{U}&=& \begin{pmatrix}  \frac{\delta b}{b} & 0 \\ 0  & 
\frac{\delta b}{b} \end{pmatrix} \\
A_x &=& A_y =0 \\
\Phi &=& 2 \sqrt{2} \alpha^{bare} \frac{\delta b}{b}  
\end{eqnarray}
The structure obtained with such biaxial uniform strain is already relaxed. The 
relaxation process is thus irrelevant for $\alpha^{bare}$, as was the use of 
canonical or DFT phonon modes for $\alpha(q)$.
We obtain the value reported in the second column of Table \ref{EPC_param}. 
Using this value and the analytical 2D 
static dielectric function $\epsilon(q)$, we can evaluate an order of magnitude 
for $\alpha(q)$ in single layer graphene. 
In the less screened case of 
suspended graphene and for $q<2k_F$, the dielectric constant is given 
by\cite{Ando2006,Hwang2008}:
\begin{eqnarray}
\epsilon(q) &=& 1+\frac{2\pi e^2}{\pi \hbar v_F} \frac{2 k_F}{|\mathbf{q}|} 
\end{eqnarray}
We find that $\alpha(q) \le \alpha(q=2k_F) \approx 0.5$ eV. 
\footnote{ Note that the DFPT values of $\alpha(q)$ are overall much 
smaller than this estimation. 
However, this is expected since $\frac{\alpha^{bare} }{\epsilon(q)}$  
refers to an isolated 
single layer of graphene whereas our DFPT calculations are performed in a 
multilayered system where the graphene planes are separated by $20$ \AA.
For the phonon momenta considered in the paper, the screening from the periodic 
images is not negligible at all.
Indeed, a phonon of wavevector $\mathbf{q}$ induces a charge fluctuation with a 
periodicity equal to $2 \pi / |\mathbf{q}|$.
The electric field induced by such a charge fluctuation decays exponentially, 
along the out-of-plane direction, on a typical length scale of 
$1/|\mathbf{q}|$. Therefore, for the interlayer distance of $20$ \AA \ used in 
our calculations, the electric field induced by the periodic images is 
negligible only for $|\mathbf{q}|$ much larger than 
$1/20$ \AA$^{-1}$ $=0.05$ \AA$^{-1}$. 
Such requirement is not satisfied by our DFPT calculations where $|\mathbf{q}|$ 
is in the range $0.013-0.077$ \AA$^{-1}$. The deformation potential is thus 
over-screened in our DFPT calculations with respect to the isolated single 
layer case.}
The presence of a substrate further enhances the screening, reducing this value.
Furthermore, as will be seen in Sec. 
\ref{sec:App_sol}, the relevant quantity to consider for resistivity is the 
squared ratio of the EPC parameter and sound velocity. The deformation 
potential term appearing only in the coupling to $\rm{LA}$ phonons while the 
gauge field term $\beta_A$ appears in the coupling to both $\rm{LA}$ and 
$\rm{TA}$, we have 
$(\frac{\alpha(q)}{v_{\rm{LA}}})^2 \approx  5\times 10^{-4} \ll 
\frac{\beta_A^2}{v_{\rm{LA}}^2+v_{\rm{TA}}^2}\approx 0.14$. This is enough to 
exclude this contribution from the following transport model. 
In the following, the deformation potential part of EPC will be ignored.
In our model, the scattering of electrons by acoustic phonons comes exclusively 
from the gauge field terms.  

\subsubsection{Calculation of strain-induced gauge-fields}

The parameter $\beta_A$ is obtained by applying a strain in the armchair 
($\mathbf{y}$) direction:
\begin{eqnarray}
\mathcal{U} &=& \begin{pmatrix}  0 & 0 \\ 0  & u_{yy} \end{pmatrix}\\
A_x &=& -\sqrt{2} \beta_A u_{yy} \\
A_y &=& \Phi =0
\end{eqnarray}
Such uni-axial strain induces a shift in the position of the Dirac point in the 
$\mathbf{x}$-direction or, equivalently, the opening of a gap $\Delta E_g$ at 
special point $\mathbf{K}$.
At this point, the value of the Hamiltonian expanded around $\mathbf{K}$ is :
\begin{eqnarray}
\mathcal{H}^{\mathcal{U}}_\mathbf{K}(\mathbf{0}) =\begin{pmatrix}  0 & 
-\sqrt{2}\beta_A u_{yy}  \\  -\sqrt{2}\beta_A u_{yy}  &  0  \end{pmatrix}  
\end{eqnarray}
The gap is thus:
\begin{eqnarray}
\Delta E_g (\mathbf{0}) &=& 2 \sqrt{2}  \beta_A u_{yy}
\end{eqnarray}
After imposing a strain on the lattice vectors, the internal coordinates of 
atoms are relaxed within DFT, and the band energies are calculated at special 
point $\mathbf{K}$. 
We repeat the process for two values of strain $ u_{yy}=-0.01, 0.01$. The 
resulting $\beta_A$ is reported in the second column of Table \ref{EPC_param}.

We repeat the whole process without relaxing the internal coordinates of the 
atoms and find the value of $\tilde{\beta_A}$ reported in Table \ref{EPC_param}.

The unscreened gauge field parameters $\tilde{\beta_A}$ and $\beta_A$ are in 
good agreement with the results of the previous section (see Table 
\ref{EPC_param} and Fig. \ref{fig:beta}), confirming the validity of the 
$|\mathbf{q}| \to 0$ and $|\mathbf{q}| = 0$ models in our simulation framework. 
The significant difference between the values of the two parameters emphasizes 
the necessity of DFT phonons and relaxation.
The strain method we propose here for acoustic parameters is especially 
well-suited for GW calculations since the energy 
bands are needed only for one $\mathbf{k}$-point.

\subsection{EPC parameters at the GW level}
\label{sec:EPC-GW}

Here we discuss how the relevant quantities are renormalized by 
electron-electron interactions within GW.
The Fermi velocity is renormalized by approximately $20\%$, depending slightly 
on doping\cite{Attaccalite2010}. The renormalization is strongest for neutral 
(or very low-doped \footnote{Strictly speaking, for graphene very close to 
charge neutrality, there is a logarithmic singularity in the group velocity 
near the Dirac point. We do not consider this effect here. By neutral graphene, 
we mean doping levels very low compared to those mentionned in the text, but 
high enough that the logarithmic singularity can be neglected}) graphene 
($\approx + 21 \%$) and slightly decreases with increasing doping 
($\approx + 17\%$ at $\varepsilon_F= 1$eV). 
It can be argued theoretically\cite{Lazzeri2008,PhysRevB.77.041409} that the 
renormalization of the coupling with phonon modes at $\mathbf{\Gamma}$ scales 
with that of the Fermi velocity because those intra-valley scattering modes 
involve only a gauge transformation (change in the position of the Dirac cone). 
This can be illustrated here by noticing that we have for acoustic phonons :
\begin{equation}
|\mathbf{\bar{K}}-\mathbf{K}|=\left| \frac{\mathbf{A}}{\hbar v_F} \right| 
\propto \frac{\beta_A}{\hbar v_F}
\end{equation}
The quantity $|\mathbf{\bar{K}}-\mathbf{K}|$ and the ratio 
$\frac{\beta_A}{\hbar v_F}$ are unaffected by electron-electron interactions 
between low energy Dirac electrons. Indeed, since such interactions are 
centro-symmetric, their inclusion cannot displace the position of the Dirac 
cone, both in presence and absence of a strain distortion.
Using the frozen phonons method, it was verified\cite{Attaccalite2010} that the 
renormalization of optical modes at $\mathbf{\Gamma}$ 
is relatively weak and is equal to the renormalization of the Fermi velocity. 
In order to verify that it is the case for acoustic modes as well,
we repeat the process of the previous paragraph within GW. The band energies 
are calculated at one additional $\mathbf{k}$-point to access the Fermi 
velocity. We use two different doping levels 
($\varepsilon_F=0.5$ eV and $0.75$ eV) to study the doping-dependency of the 
renormalization. The doping levels are rather high to ensure that the Fermi 
surface is satisfactorily sampled by our grid.
 The results are presented in Table \ref{table:GW}.
\begin{table}
\caption{Renormalization of Fermi velocity and acoustic gauge field parameter 
by electron-electron interactions within GW, presented for two different values 
of doping and for uniaxial strain $u_{yy}=-0.01, +0.01$ as explained in Sec. 
\ref{sec:strain-Num}. }
\begin{center}
\begin{tabular*}{0.43\textwidth}{ l | c | c | c | c  }
\hline 
\hline
Strain $u_{yy}$  & \multicolumn{2}{c|}{$-0.01$}& \multicolumn{2}{c}{$+0.01$}  \\

Fermi energy $\varepsilon_F$ & $0.50$ eV &$0.75$ eV &$0.50$ eV &$0.75$ eV \\
$v_F^{GW}/v_F^{DFT}$    & $1.203$ &  $1.166$    & $1.204$  & $1.164$  \\
$\beta_A^{GW}/\beta_A^{DFT}$ & $1.215$  &  $1.172$    & $1.202$   & $1.165$ \\
\end{tabular*}
\end{center}
\label{table:GW}
\end{table} 

Our calculations confirm the $\approx 20 \%$ renormalization of $v_F$ at low 
doping. More importantly, they confirm what was assumed in our previous 
work\cite{Park2014}, namely that the electron-electron interactions renormalize 
$\beta_A$ as the Fermi velocity:
$$\frac{v_F^{GW}}{v_F^{DFT}} \approx \frac{\beta_A^{GW}}{\beta_A^{DFT}}$$
In contrast, the interaction of electrons with inter-valley A$_1'$ mode is not 
just a gauge transformation of the electronic Hamiltonian.
The renormalization of this mode is much stronger overall, and its doping 
dependency is more pronounced.
According to Ref. \onlinecite{Attaccalite2010}, $\beta_K$ is renormalized by 
$\approx +46\%$ close to neutrality, and $\approx 20 \%$ at 
$\varepsilon_F= 1$eV. 

As will be discussed in more details in the rest of this paper, the 
contribution to the resistivity of each phonon modes is proportional to the 
squared ratio of the EPC parameter and the Fermi velocity. For the phonon modes 
at $\mathbf{\Gamma}$, the Fermi velocity renormalization is exactly compensated 
by the electron-phonon coupling renormalization. The GW corrections thus has no 
effect on their contribution resistivity. 
The A$_1'$ mode, on the other hand, is renormalized more strongly than the 
Fermi velocity. Therefore, we have to choose relevant values of the 
renormalization for both the Fermi velocity and $\beta_K$. In Sec. 
\ref{sec:Results}, we will see that the available experimental data allows for 
a comparison of the value of $\beta_K$ only in a short range of doping close to 
neutrality ($0.12 \to 0.21$eV). 
 
In view of the above remarks, we use in our resistivity simulations the
 $+ 46 \%$ renormalization of $\beta_K$ obtained at neutrality by Ref. 
\onlinecite{Attaccalite2010}. We then choose the corresponding $+ 20 \%$ 
renormalization of the Fermi velocity, relevant at low doping.
Finally, the coupling to phonons modes at $\mathbf{\Gamma}$ ($\beta_A$, 
$\beta_O$) are renormalized by $+ 20 \%$, as the Fermi velocity. 

\section{Boltzmann transport theory}
\label{sec:BTT}

In this section we present a numerical solution to the Boltzmann transport 
equation for phonon-limited transport in graphene. The general method presented 
here is well known in carrier transport theory and has been applied to some 
extent to graphene\cite{Suzuura2002,Shishir2009,
Perebeinos2010,Mariani2010,Hwang2008a,Hwang2007}. The central addition to those 
previous works resides in the treatment of a more complete EPC model, and a 
numerical solution involving very few approximations. The initial steps are 
repeated in an effort to clarify the assumptions involved. 

A carrier current is created by applying an electric field $\mathbf{E}$. This 
has the effect of changing the electronic distribution $f(\mathbf{k})$. In the 
steady state regime, the new distribution favors states with electron momentum 
$\mathbf{k}$ in the opposite direction of the electric field, thus creating a 
net current $\mathbf{j}$.
We are interested in the current in the direction of the electric field that we 
choose to be the $\mathbf{x}$-axis, used as a reference for angles in our 
model. Throughout this work, the resistivity is to be understood as the 
diagonal part of the resistivity tensor $\rho=\rho_{xx}$. It is given 
by\cite{Ashcroft1976}:
\begin{equation}
\frac{1}{\rho}= \frac{\mathbf{j} \cdot \mathbf{x}}{|\mathbf{E}|} =
 \frac{2 e}{|\mathbf{E}|} \int_{BZ} \frac{d\mathbf{k}}{(2 \pi)^2} 
 f(\mathbf{k}) \mathbf{v(k)}   \cdot \mathbf{x}
\label{eq:current}
\end{equation}
The integral is made over the Brillouin zone with a factor $2$ for spin 
degeneracy, $e$ is the elementary charge, $\mathbf{v(k)}$ is the carrier 
velocity of state $|\mathbf{k}\rangle$, and $\mathbf{x}$ is the Cartesian 
coordinate basis vector. 
In the framework of linear response theory, we are interested in the response 
of $f(\mathbf{k})$ to the first order in electric 
field\cite{Nag1980,Sarma1992,Basu1980}. Consistent with the Dirac cone model, 
we assume $\varepsilon_{\mathbf{k}}=\hbar v_F |\mathbf{k}|$ and 
$\mathbf{v(k)}= v_F \frac{\mathbf{k}}{|\mathbf{k}|}$. We then separate the norm 
and angular dependency of $\mathbf{k}$ in 
$f(\mathbf{k})=f(\varepsilon_{\mathbf{k}}, \theta_{\mathbf{k}})$ and consider 
the first order expansion:
\begin{equation}
f(\varepsilon_{\mathbf{k}},\theta_{\mathbf{k}})= 
f^{(0)}(\varepsilon_{\mathbf{k}})+
f^{(1)}(\varepsilon_{\mathbf{k}}, \theta_{\mathbf{k}})
\label{devf}
\end{equation}
where $ f^{(1)}$ is proportional to the electric field, and $f^{(0)}$ is the 
equilibrium Fermi-Dirac distribution, which has no angular dependency. Due to 
graphene symmetries, the two Dirac cones of the Brillouin zone give the same 
contribution to Eq. \ref{eq:current}.
Multiplying by a factor 2 for valley degeneracy, performing the dot product, 
and using $f$, Eq. \ref{eq:current} becomes:
\begin{equation}
\frac{1}{\rho} = \frac {4 e}{|\mathbf{E}|} \int_{\mathbf{K}} 
\frac{d\mathbf{k}}{(2 \pi)^2} 
f^{(1)}(\varepsilon_{\mathbf{k}}, \theta_{\mathbf{k}}) v_F 
\cos(\theta_{\mathbf{k}}) 
\label{eq:rho}
\end{equation}
where the integral is now carried out within a circular region around 
$\mathbf{K}$. It is clear that the zeroth order term gives no contribution due 
to the angular integral, and that we have to look for 
$f^{(1)} \propto \cos(\theta_{\mathbf{k} } )$. 
\\

We now use Boltzmann transport equation to obtain the energy dependency of 
$f^{(1)} $. A key quantity is the collision integral 
$ \left( \frac{\partial f}{\partial t} \right)_{coll} (\mathbf{k})$
describing the rate of change in the occupation of the electronic state  
$|\mathbf{k} \rangle$ due to scattering. Assuming that the electronic 
distribution is in a spatially uniform, out-of-equilibrium but steady state, 
the change of the occupation function triggered by the electric field must be 
compensated by the change due to collisions\cite{Ashcroft1976}:
\begin{equation}
- \frac{e\mathbf{E}}{\hbar} \cdot \frac{\partial f}{\partial \mathbf{k}} =
\left( \frac{\partial f}{\partial t} \right)_{coll} (\mathbf{k}) \label{BTE}  \\
\end{equation}
Using Fermi golden rule, the collision integral is:
\begin{eqnarray}
&&\left( \frac{\partial f}{\partial t} \right)_{coll} (\mathbf{k}) =  
\label{coll_int} \\
&&\sum_{\mathbf{k'}} \left\{P_{\mathbf{k'}\mathbf{k}}  
f(\mathbf{k'})(1-f(\mathbf{k}))- P_{\mathbf{k}\mathbf{k'}} 
f(\mathbf{k})(1-f(\mathbf{k'})) \right\} \nonumber
\end{eqnarray}
Here, $\mathbf{k}$ belongs to a circular region around $\mathbf{K}$. The 
momentum of the scattered states $\mathbf{k'}$ is i) around $\mathbf{K}$ for 
intra-valley scattering modes; ii) around the other Dirac cone $\mathbf{K'}$ in 
case of inter-valley scattering. The quantity $P_{\mathbf{k}\mathbf{k'}}$ is 
the scattering probability from state $|\mathbf{k}\rangle$ to 
$|\mathbf{k'}\rangle$.
It satisfies the detailed balance condition, namely:
\begin{eqnarray}
P_{\mathbf{k'}\mathbf{k}}  f^{(0)}(\varepsilon_{\mathbf{k'}}) 
(1- f^{(0)}(\varepsilon_{\mathbf{k}}))=  P_{\mathbf{k}\mathbf{k'}}  
f^{(0)}(\varepsilon_{\mathbf{k}}) 
(1- f^{(0)}(\varepsilon_{\mathbf{k'}}))\nonumber\\
\label{eq:detailed}
\end{eqnarray}
The scattering probability $P_{\mathbf{k}\mathbf{k'}}$
is composed of two terms,
\begin{eqnarray}
P_{\mathbf{k}\mathbf{k'}} = P_{\mathbf{k}\mathbf{k'}, \rm{I} } +
 \sum_{\nu} P_{\mathbf{k}\mathbf{k'}, \nu} 
\label{eq:additivity}
\end{eqnarray}
where $P_{\mathbf{k}\mathbf{k'}, \rm{I}}$ is the impurity scattering probability
and $P_{\mathbf{k}\mathbf{k'}, \nu}$ is due to the electron-phonon scattering 
of the $\nu^{\rm th}$ phonon branch. 

In the Born approximation, the impurity scattering probability is
\begin{eqnarray}
P_{\mathbf{k}\mathbf{k'}, \rm{I}} &=& \frac{2\pi}{\hbar} \frac{1}{N}  
n_i |\langle \mathbf{k} |H_i | \mathbf{k'} \rangle |^2 
\delta(\varepsilon_{\mathbf{k}}-\varepsilon_{\mathbf{k'}})
\label{eq:Pimp}
\end{eqnarray}
where $H_i$ is the electron-impurity interaction Hamiltonian, $n_i$ the 
impurity density and $N$ is the number of unit cells. When needed for 
comparison with experiment, we use existing 
methods\cite{Hwang2007,Hwang2008b,Barreiro2009} to fit charged and short-range 
impurity densities on the low temperature resistivity measurements.
The electron-phonon scattering probability is given by 
\begin{eqnarray}
P_{\mathbf{k},\mathbf{k+q}, \nu} &=& \frac{2\pi}{\hbar} \frac{1}{N}  
|g_{\mathbf{k+q},\mathbf{k},\nu}|^2  
\left\{n_{q,\nu}\delta(\varepsilon_{\mathbf{k+q}}-
\varepsilon_{\mathbf{k}}-\hbar\omega_{\mathbf{q},\nu} )  \right. \nonumber\\
&+&\left.(n_{q,\nu}+1)\delta(\varepsilon_{\mathbf{k+q}}-\varepsilon_{\mathbf{k}}
+\hbar \omega_{\mathbf{q},\nu}) \right\} 
\label{eq:Pep}
\end{eqnarray}
where $\mathbf{k+q}=\mathbf{k'}$, $g_{\mathbf{k+q},\mathbf{k},\nu}$ is the 
electron-phonon matrix element introduced in 
Eq. \ref{eq:defg} and $n_{q,\nu}$ is the Bose-Einstein equilibrium occupation 
of mode $\nu$ with phonon-momentum ${\bf q}$ and phonon frequency 
$\omega_{\mathbf{q},\nu}$.

By Replacing Eq. \ref{devf} in Eq. \ref{coll_int}, using Eq. \ref{eq:detailed} 
and keeping only first order terms we obtain
\begin{widetext}
\begin{equation}
-\frac{e\mathbf{E}}{\hbar} \cdot \frac{\partial f}{\partial \mathbf{k}}  =   
\sum_{\mathbf{k'}} P_{\mathbf{k}\mathbf{k'}} 
\frac{1-f^{(0)}(\varepsilon_{\mathbf{k'}})}{1- 
f^{(0)}(\varepsilon_{\mathbf{k}})}
\Big(  f^{(1)}(\varepsilon_{\mathbf{k'}}, \theta_{\mathbf{k'}}) 
\frac{f^{(0)}(\varepsilon_{\mathbf{k}}) 
(1- f^{(0)}(\varepsilon_{\mathbf{k}}))  }
{f^{(0)}(\varepsilon_{\mathbf{k'}})(1- f^{(0)}(\varepsilon_{\mathbf{k'}}))} -
f^{(1)}(\varepsilon_{\mathbf{k}}, \theta_{\mathbf{k}})  \Big)
\label{eq:big1}
\end{equation}
\end{widetext}

We first consider the left-hand side of Eq. \ref{eq:big1} and, looking for a 
solution that is first order in the electric field, we have:
\begin{eqnarray}
-\frac{e\mathbf{E}}{\hbar} \cdot \frac{\partial f}{\partial \mathbf{k}}  &=&  - 
F \cos(\theta_{\mathbf{k}})   \hbar v_F 
\frac{\partial f^{(0)}}{\partial \varepsilon} (\varepsilon_{\mathbf{k}})  
\label{eq:lhsbig1}
\end{eqnarray}
with $F =   \frac{e|\mathbf{E}|}{\hbar}$. In the right-hand-side of Eq.  
\ref{eq:big1}  we adopt\cite{Sarma1992} the following ansatz for
 $f^{(1)}(\varepsilon_{\mathbf{k}}, \theta_{\mathbf{k}})$:
\begin{equation}
f^{(1)}(\varepsilon_{\mathbf{k}}, \theta_{\mathbf{k}}) =F 
\tau(\varepsilon_{\mathbf{k}})\cos(\theta_{\mathbf{k}})   \hbar v_F 
\frac{\partial f^{(0)}}{\partial \varepsilon} (\varepsilon_{\mathbf{k}})
\label{eq:rhsbig1}
\end{equation}
and verify that it solves Eq. \ref{eq:big1}. The energy dependency of $f^{(1)}$ 
is captured by $\tau(\varepsilon_{\mathbf{k}})$. 
This auxiliary variable has the dimension of time. The time 
$\tau(\varepsilon_{\mathbf{k}})$ depends on the perturbation and has a meaning 
only in the the framework studied here, namely the steady state of a 
distribution under an external electric field. It is different from the 
relaxation time from relaxation time approximation and from the scattering time 
as can be measured by angle resolved photoemission spectroscopy.

By replacing Eqs. \ref{eq:lhsbig1} , \ref{eq:rhsbig1} in Eq. \ref{eq:big1} we 
obtain the following equation, as in Refs. 
\onlinecite{Nag1980,Sarma1992,Basu1980}, and dividing both members by 
$\cos(\theta_{\mathbf{k}})$
we obtain:
\begin{eqnarray}
1   & =&   \sum_{\mathbf{k'}} P_{\mathbf{k}\mathbf{k'}}  
\frac{  1- f^{(0)}(\varepsilon_{\mathbf{k'}})}{ 1- 
f^{(0)}(\varepsilon_{\mathbf{k}})} 
\left( \tau(\varepsilon_{\mathbf{k}})      - \tau(\varepsilon_{\mathbf{k'}}) 
\frac{ \cos(\theta_{\mathbf{k'}}) }{ \cos(\theta_{\mathbf{k}})}  \right) 
\nonumber \\
\label{eq:BT1}
\end{eqnarray}

We then parametrize $\mathbf{k}$-space in energy (equivalent to norm through 
$\varepsilon= \hbar v_F |\mathbf{k}|$). On an energy grid of step 
$\Delta \varepsilon$, Eq. \ref{eq:BT1} can be written: 
$$\sum_{\varepsilon'}\mathcal{M}_{\varepsilon,\varepsilon'}\tau(\varepsilon') 
=1 $$
where $\varepsilon=\hbar v_F |\mathbf{k}|$, $\varepsilon'=\hbar v_F 
|\mathbf{k'}|$ and the matrix $\mathcal{M} $ is defined in appendix  \ref{Num}.
The time $\tau$ can then be obtained by  numerical inversion of the matrix  
$\mathcal{M}$ (see App. \ref{Num} for more details).

It is worthwhile to recall that, due to the additivity of the scattering 
probabilities in Eq.  \ref{eq:additivity}, the 
matrix $\mathcal{M}$ involves a sum over impurities and different phonon bands, 
namely $\mathcal{M}=\mathcal{M}_{\rm{I}} + \sum_{\nu}\mathcal{M}_{\nu} $.
Strictly speaking, the time $\tau$ is obtained from the inversion 
of the global matrix $\mathcal{M}$ and not from the sum of the inverse of the 
matrices $\mathcal{M}_{\nu}$ and $\mathcal{M}_{\rm{I}}$. The latter is 
equivalent to applying Matthiessen's rule, which is an approximation (see Sec. 
\ref{sec:App_sol}).
 
From the time $\tau$ we obtain the distribution function. Inserting it into Eq. 
\ref{eq:current}, the electrical resistivity is found by evaluating the 
following integral numerically:
\begin{align}
\frac{1}{\rho}&=\frac{e^2 v_F^2}{2} \int  d\varepsilon DOS(\varepsilon)    
\tau(\varepsilon)
\left( -\frac{\partial f^{(0)}}{\partial \varepsilon} (\varepsilon)  \right) 
\label{eq:resistivity}
\end{align}

Where $DOS(\varepsilon_{\mathbf{k}})=
\frac{ 2 |\varepsilon_{\mathbf{k}}|}{ \pi (\hbar v_F)^2}$ is the total density 
of states per unit area of graphene, valley and spin degeneracy included.

\section{Results}
\label{sec:Results}

\begin{table}[h]
\begin{tabular}{ l c c c }
\hline
 Experiment & I & II  \\
\hline 
Temp. range (K) &   $4 \sim 250$  &  $  14 \sim 480 $\\
Doping range (eV)  &  $0.36 \sim 1.01$  & $0.09 \sim 0.21$   \\
Gate dielectric & PEO   &  SiO2   \\ \hline
\end{tabular}
\caption{Sources of the experimental data used in this section. Experiment I 
and II correspond to Ref. \onlinecite{Efetov2010} and Ref. 
\onlinecite{Chen2008} respectively. Combining those two experiments provides us 
with  a wide range of temperature and doping conditions. PEO stands for 
poly(ethylene)oxide.}
\label{tab:Exp_sources}
\end{table}

\begin{table}[h]
\caption{Numerical values of parameters used in resistivity calculations. The 
effective sound velocity $v_{\rm{A}}$ is defined in Eq. \ref{eq:va}.  }
\begin{tabular}{ l c c c }
\hline
Parameter & symbol  &Value  \\
\hline 
Acoustic gauge field (GW) & $\beta_A$ & $4.32$ eV  \\
Acoustic gauge field (fitted)& $\beta_A$ & $4.97$ eV  \\
Optical gauge field (GW) & $\beta_O$  & $11.4$ eV/\AA  \\
A$_1'$ EPC parameter (GW) & $\beta_K$  & $13.9$ eV/\AA  \\ 
A$_1'$ EPC parameter (fitted) & $\beta_K$  & see Fig. \ref{fig:betaK}  \\ 
Lattice parameter & $a$ & $2.46$ \AA \\
Unit-cell area & $S_{\Re}$ & $5.24$ \AA$^2$ \\
Sound velocity TA & $v_{\rm{TA}} $ & $13.6$ km.s$^{-1}$ \\
Sound velocity LA & $v_{\rm{LA}} $ & $21.4$ km.s$^{-1}$ \\
Effective sound velocity & $v_{\rm{A}} $ & $16.23$ km.s$^{-1}$ \\
LO/TO phonon energy & $\hbar \omega_{\rm{LO}/\rm{TO}}$& $0.20$ eV  \\
A$_1'$ phonon energy & $\hbar \omega_{\rm{A}_1'}$& $0.15$ eV  \\
Carbon atom mass &  $M$ & $12.0107$ u   \\
Mass density     & $\mu_S=2M/S_{\Re}$ &  $7.66$ kg/m$^2$    \\
Fermi Velocity (GW) & $v_F$ & $1.00$ $10^6$ ms$^{-1}$ \\
 \hline
\end{tabular}
\label{tab:Sim_param}
\end{table}

\begin{figure*}[t]
\subfigure[]{\includegraphics{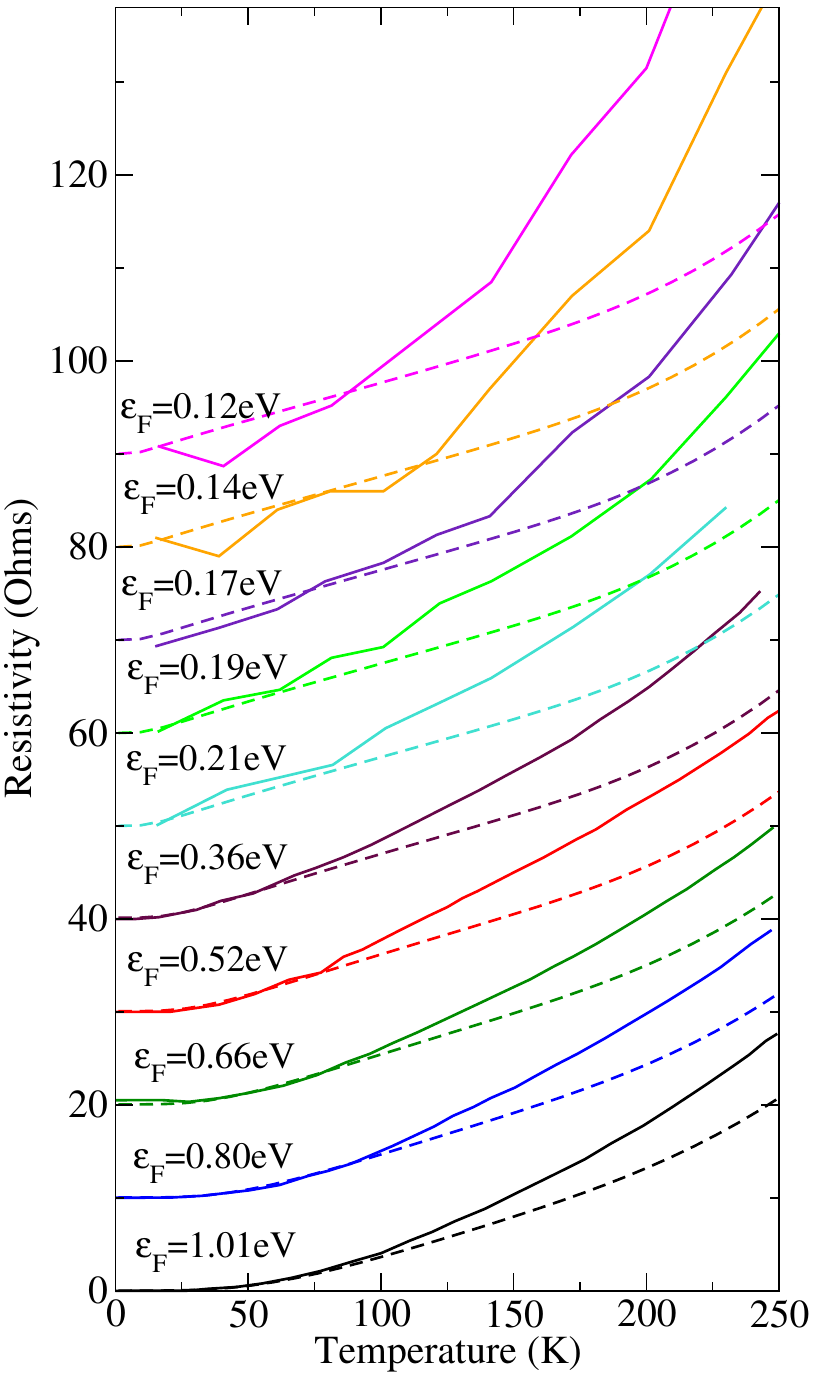} \label{fig:Acoustic_all}}
\subfigure[]{\includegraphics{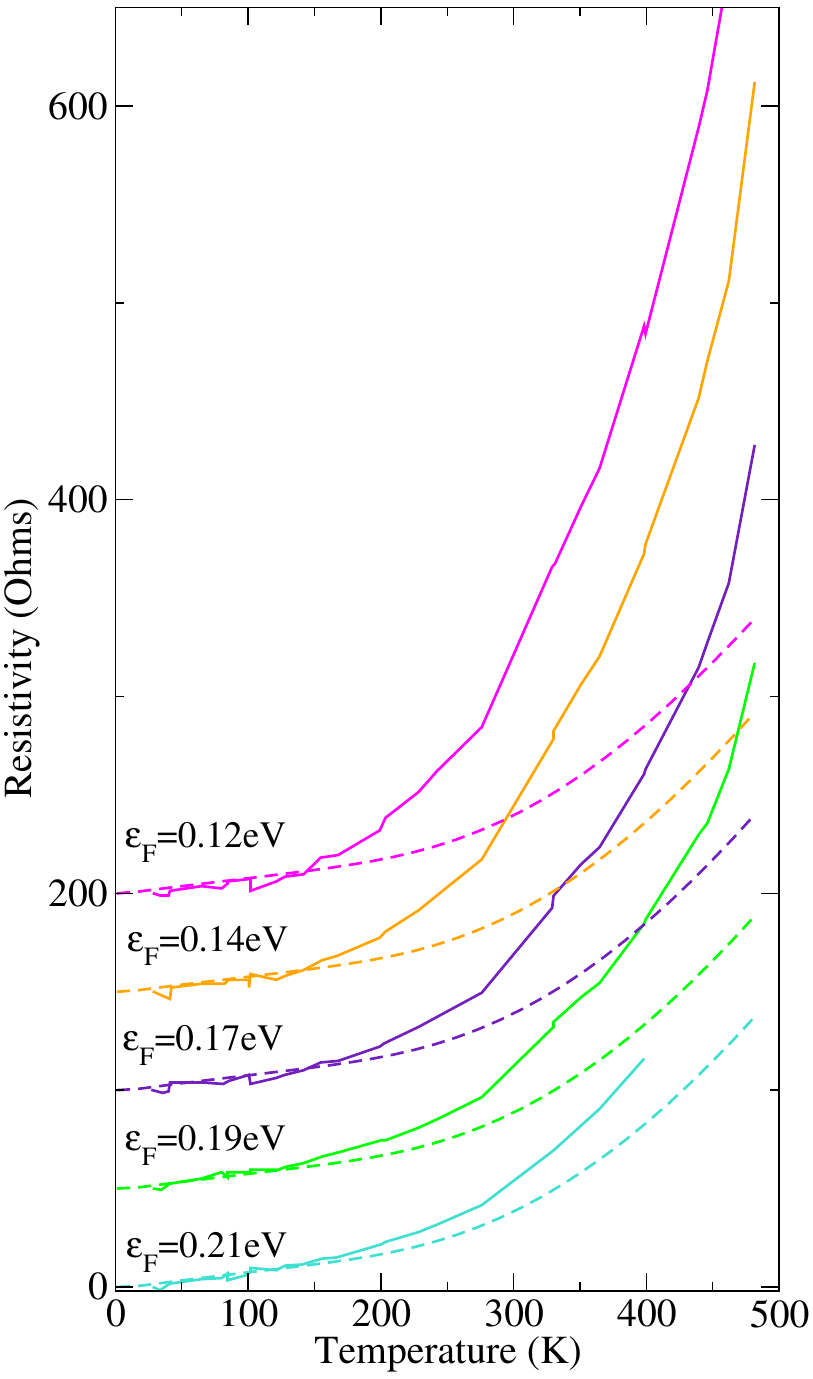}\label{fig:Optics}}
\caption{(Color online) Comparison of experimental data (plain lines) and the 
numerical solution of the Boltzmann equation (dashed lines) in the 
Bloch-Gr\"uneisen and
Equipartition regimes(a) and in the HT regime(b), for several doping levels 
ranging from $\varepsilon_F=0.12$ eV to $\varepsilon_F=1.01$ eV. Experimental 
data are from Refs. \onlinecite{Chen2008} ($0.12\to 0.21$eV) and 
\onlinecite{Efetov2010} ($0.36\to 1.01$eV). For each curve, the residual 
resistivity was subtracted, then a fictitious residual resistivity was added 
for clarity of the plot (different values were used in the two plots). Though 
their value has no absolute sense, the fictitious resistivities are ordered as 
the real ones.}
\label{fig:Num-Exp-unfit}
\end{figure*}

\begin{figure*}[t]
\subfigure[]{\includegraphics{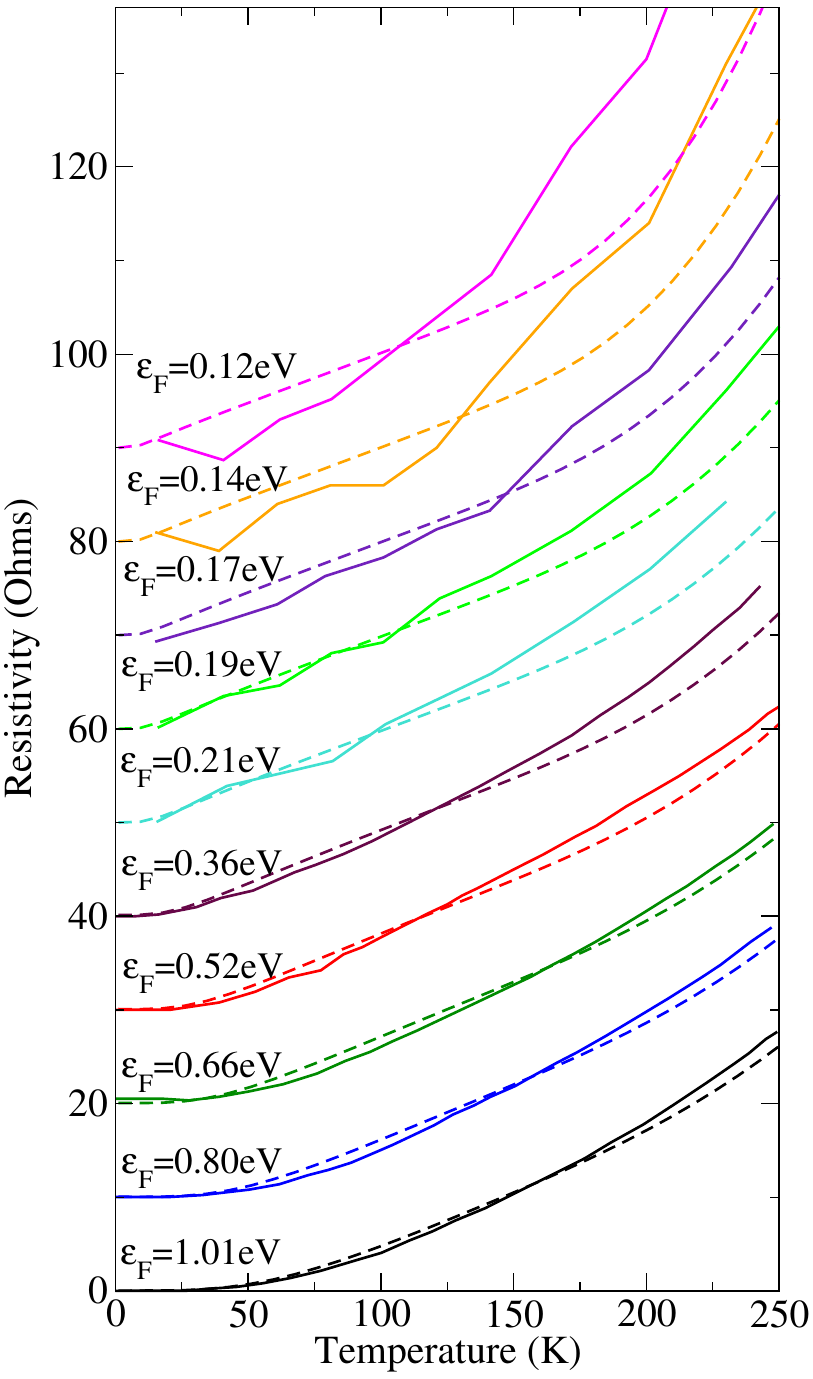}\label{fig:Acoustic_all_fit}}
\subfigure[]{\includegraphics{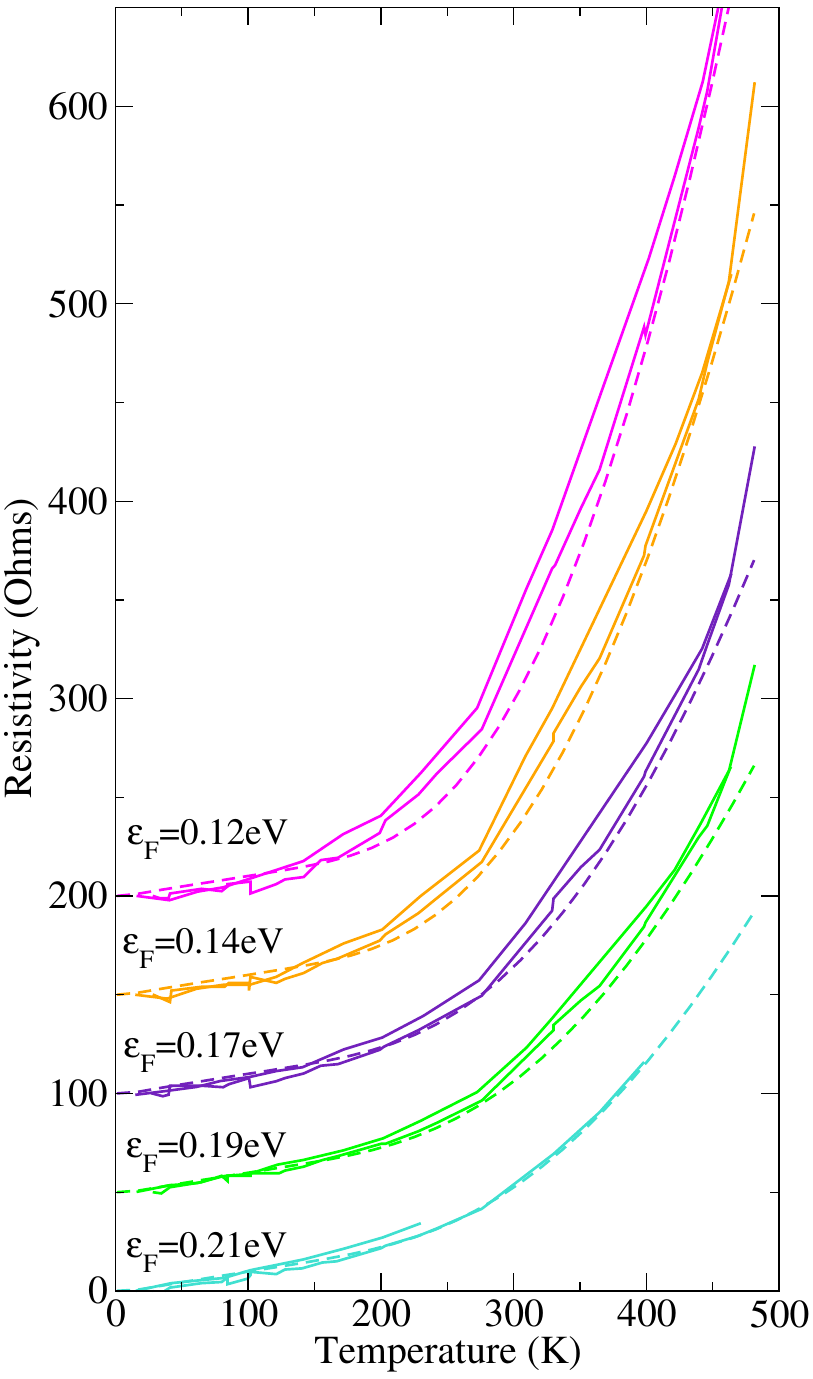}\label{Optics_fit}}
\caption{(Color online) Results presented as in Fig. \ref{fig:Num-Exp-unfit}, 
with fitted values for $\beta_A$ (see Table \ref{EPC_param}) and $\beta_K$ (see 
Fig. \ref{fig:betaK}) in the simulations.
Experimental data from a different sample are added in plain lines in Fig. 
\ref{Optics_fit} to show that the sample-to-sample discrepancy when approaching 
the Dirac point is of the same order as the disagreement with the fitted 
numerical model. }
\label{fig:Num-Exp-fit}
\end{figure*}

On general grounds, three different regimes are present in our calculations. 
These three regimes depend on three energy scales:
i) the Fermi energy $\varepsilon_F$ is the reference energy around which 
initial and scattered states are situated; ii) the phonon energy $\hbar \omega$ 
is the energy difference between initial and scattered states; iii) the  
temperature $k_BT$ sets the interval on which electronic and phonon occupations 
vary. Comparing $\hbar \omega$ to $\varepsilon_F$ indicates how much the 
density of states changes during a scattering process. Comparing $\hbar \omega$ 
to $k_BT$ indicates phonon occupations and the change of electronic 
occupations. Based on those observations, we have:
\begin{itemize}
\item {\it Bloch-Gr\"uneisen} (BG) regime ($0K<T \lesssim 0.15 \times T_{BG}$ 
where $T_{BG}= 2\hbar v_{TA/LA} k_F/k_B$). At these temperatures, $k_BT$ is 
small compared to the energy of optical phonons. Those
modes are not occupied and do not contribute. In contrast, the acoustic modes 
contribute since $k_B T$ is of the order of 
$\hbar \omega_{\mathbf{q},\rm{TA}/\rm{LA}}$. Moreover,  
the occupancy of initial states $f(\varepsilon_{\mathbf{k}})$ and scattered 
states 
$f(\varepsilon_{\mathbf{k}}\pm \hbar \omega_{\mathbf{q},\rm{TA}/\rm{LA}})$ are 
significantly different. Finally,
as $ \hbar \omega_{\mathbf{q},\rm{TA}/\rm{LA}}<< \varepsilon_F$, quantities 
other than occupancy, such as the density of states, can be considered 
constant. 
In this regime, the resistivity has a $\propto T^4$ dependence due to acoustic 
phonons.

\item {\it Equipartition} (EP) regime ($0.15 \times T_{BG} \lesssim T \lesssim 
0.15 \times \hbar \omega_{A_1'}\approx 270$K): optical phonons still do not 
contribute.
As $ \hbar \omega_{\mathbf{q},\rm{TA}/\rm{LA}} << k_BT<< \varepsilon_F $ 
the scattering by acoustic phonons can be approximated as elastic, including in 
the occupancies of initial and final states. 
The resulting resistivity is then linear in temperature.

\item {\it High temperature} (HT) regime 
($T \gtrsim 0.15 \times \hbar \omega_{A_1'}\approx  270$K) : the elastic 
approximation for acoustic phonons holds, 
but the three energy scales are comparable in the case of optical phonons. 
In this case, no reasonable approximation can be made globally. 
The optical phonon participation is characterized by a strongly increasing 
resistivity at a temperature around $15\%$ of the phonon energy. Due to their 
lower energy and stronger coupling, the contribution of optical A$_1'$ phonons 
is more pronounced than LO/TO phonons.
\end{itemize}

Our calculated resistivity is compared with experimental data in Fig. 
\ref{fig:Num-Exp-unfit}.
Experimental data  are from the references cited in Table \ref{tab:Exp_sources} 
and the computational parameters are summarized in Table \ref{tab:Sim_param}.
\\

Below room temperature, Fig. \ref{fig:Acoustic_all}, the comparison between 
theory and experiment is meaningful only above $\varepsilon_F \gtrsim 0.14$eV.
Indeed, when approaching the Dirac point\cite{Hwang}, the 
electron density tends to zero and resistivity diverges.
One has to adopt a model with non-homogenous electron 
density\cite{Martin2007} to obtain a finite resistivity, such that the Fermi 
energy is ill-defined. Temperature-dependent screening of impurity scattering 
as well as temperature-dependent chemical potential shift\cite{Hwang2009} also 
play a role in this regime. Those issues are not treated in our model.
At sufficiently high doping, the temperature behavior of BG and EP regimes
are well reproduced, despite an overall underestimation. 
The doping-dependency of the resistivity is limited to the BG 
($\rho\propto T^4$) regime. 
The upper boundary of this regime increases with 
doping, since $T_{BG} \propto k_F$. Above $\sim 0.15 \times T_{BG}$, in the EP 
regime, the slope of the resistivity is essentially doping independent.
This confirms that the deformation potential term can be neglected, since its 
screening would induce such a dependency.  

In the HT regime, Fig. \ref{fig:Optics}, the underestimation is globally more 
pronounced. 
The increase of experimental resistivity around room temperature is steeper 
than the theoretical one. 
A strong doping dependency of the experimental curves appear. The agreement 
with the simulations improves as the system is doped
far away from the Dirac point. 
Usually this discrepancy is attributed to remote-phonon scattering from the 
SiO$_2$ substrate\cite{Chen2008}. 
This effect is missing in our calculation as the substrate is not included.
Moreover, we found no experimental data on other substrates for those 
temperatures. 
It follows that substrate dependent sources of scattering cannot be ruled out 
in the HT regime.
However, we would like to point out that based on the observation that the 
contribution of optical phonons seems to appear at a temperature 
$\approx 0.15 \times \hbar \omega_{\nu}$, we expect intrinsic optical phonons 
to be better candidates than the relatively low energy remote-phonons proposed 
in Ref. \onlinecite{Chen2008}.
The optical A$_1'$ mode at $\mathbf{K}$ does induce a sudden increase of 
resistivity, and the temperature at which this occurs is in very good agreement 
with experiment. The increasing discrepancy between theory and experiment in 
the magnitude of resistivity at lower doping could be explained by the fact that
the EPC parameter $\beta_K$ corresponding to the A$_1'$ mode is
renormalized by electron-electron interaction\cite{Attaccalite2010}. This 
renormalization decreases the larger the electron-doping of graphene, and tends
to the DFT value at high doping. 

We then fit the value of $\beta_A$ for acoustic modes on experimental data in 
the EP regime. We find that an increase of the electron-phonon coupling of the
acoustic modes of $15\%$ leads to an excellent agreement with
experimental data in the BG and EP regimes,
as shown in Fig. \ref{fig:Acoustic_all_fit}. We found an equivalent agreement 
for resistivity measurements of graphene on h-BN\cite{Dean2010} or on SiO$_2$ 
with HfO$_2$ gate dielectric\cite{Zou2010}, thus ruling out any significant 
contribution from substrate dependent sources of scattering (other than charged 
and short-range impurities) in the BG and EP regimes.
We then conclude that the solution of the Boltzmann equation based on DFT and 
GW  (the two methods are equivalent here) electron-phonon coupling
parameters and bands explains fairly well the low-temperature regime 
(BG and EP), although DFT seems to underestimate the coupling to acoustic modes 
by $15\%$, or the resistivity by $\approx 30\%$. On closer inspection (see Sec. 
\ref{sec:App_sol} and App. \ref{relax-times}), the resistivity in the EP regime 
is proportional to $\frac{\beta^2_A}{v_{\rm{A}}^2}$ where $v_{\rm{A}}$ is the 
effective sound velocity given in Table \ref{tab:Sim_param}. An overestimation 
of $v_{\rm{A}}$ could also explain the $30\%$ underestimation of the 
resistivity. Finally, this discrepancy might be partly due to some other 
$\propto T$ contribution from processes ignored here, such as impurity 
scattering with temperature-dependent screening\cite{Hwang2009}. In any case, 
defining an effective parameter $\beta_A$ with the fitted value found here is 
sufficient to describe low temperature resistivity in a relatively large range 
of doping levels.   

\begin{figure}[h]
\includegraphics{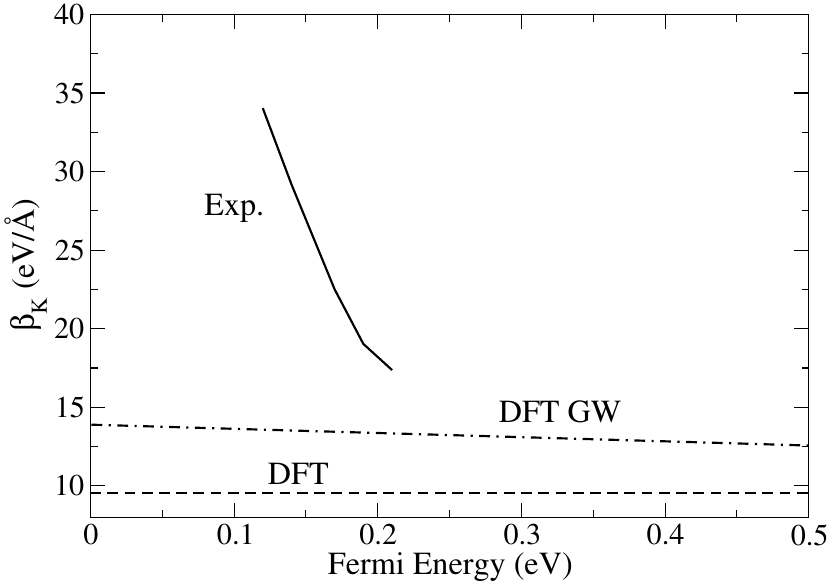}
\caption{The parameter $\beta_K$ fitted on experimental results as a function 
of Fermi energy is compared to the DFT LDA and DFT GW values. The DFT GW values 
are from Ref. \onlinecite{Attaccalite2010}.}
\label{fig:betaK}
\end{figure}

Within such a picture, we fit the optical $\beta_{K}$ parameter as a function 
of doping. The fitted values of $\beta_K$ are plotted in Fig.  \ref{fig:betaK}, 
along with GW and DFT values. Near the Dirac point, the fitted coupling 
parameter 
increases substantially more than previous estimates\cite{Attaccalite2010} at 
the GW level, 
but at high doping it seems to approach the DFT value.
We then plot the resistivity with the fitted $\beta_A$ and $\beta_K$,
and find a good agreement with experiments on Fig. \ref{Optics_fit}.
However, since the screened coupling to remote phonon has a similar behavior
as a function of doping, it is not possible to rule out this effect. 

\section{Approximated solutions}
\label{sec:App_sol}

In this section we seek a compromise between analytical simplicity and 
numerical accuracy. We review some of the approximations often made in 
transport models, and check their validity against the full numerical solution 
presented in Sec. \ref{sec:BTT}. Fitted EPC parameters are used in the 
resistivity calculations.

\subsection{Semi-analytical approximated solution}
\label{sec:semianal}

The first essential step is to derive an
analytical expression of the time
$\tau_{\nu}(\varepsilon_{\mathbf{k}})$ for each phonon branch.
We first rewrite equation \ref{eq:BT1} as:
\begin{eqnarray}
 \frac{1}{\tau(\varepsilon_{\mathbf{k}})}  & =&  \sum_{\mathbf{k'}} 
P_{\mathbf{k}\mathbf{k'}}  \frac{  1- f^{(0)}(\varepsilon_{\mathbf{k'}})}{ 1- 
f^{(0)}(\varepsilon_{\mathbf{k}})}
\nonumber  \\
&&\times   \Big( 1     - \frac{\tau(\varepsilon_{\mathbf{k'}})  
\cos(\theta_{\mathbf{k'}})}{\tau(\varepsilon_{\mathbf{k}}) 
\cos(\theta_{\mathbf{k}})}    \Big)
\label{eq:BT2b}
\end{eqnarray}
For the doping level considered here, impurity scattering is essentially 
constant on the energy scale of the phonon energies. When this type of 
scattering dominates, the approximation
$\tau(\varepsilon_{\mathbf{k'}}) \approx \tau(\varepsilon_{\mathbf{k}})$ 
becomes reasonable. We can simplify $\tau(\varepsilon_{\mathbf{k}})$ on the 
right-hand side 
of Eq. \ref{eq:BT2b}  and write:
\begin{equation}
\frac{1}{\tau(\varepsilon_{\mathbf{k}})} = \sum_{\nu} 
\frac{1}{\tau_{\nu}(\varepsilon_{\mathbf{k}})} 
\end{equation}
In other words, Matthiessen's rule\cite{Matthiessen1864} can be applied. With 
mild restrictions on the form of the angular dependency of the scattering 
probability, one can then use the following expression for the times 
$\tau_{\nu}$\cite{Kaasbjerg2012,Hwang2008a}:
\begin{equation}
\frac{1}{\tau_{\nu}(\varepsilon_{\mathbf{k}})} \approx  \sum_{\mathbf{k'}} 
P_{\mathbf{k}\mathbf{k'}, \nu} \frac{  1-
  f^{(0)}(\varepsilon_{\mathbf{k'}})}{ 1-
  f^{(0)}(\varepsilon_{\mathbf{k}})} \Big(  1  -
\cos(\theta_{\mathbf{k'}}-\theta_{\mathbf{k}}) \Big) \\
\label{eq:1otaunu}
\end{equation} 
A solution of Eq. \ref{eq:1otaunu} can now be carried out for
different times separately by using phonon-specific approximations. 
Details can be found in App. \ref{relax-times}. We present here a solution that 
is relatively simple, yet very close to the complete one in a large range of 
temperature. 
For the sum of TA and LA acoustic phonons, labeled by the index 
$\rm{A} \equiv \rm{TA}+\rm{LA}$, we use the time derived in the EP and HT 
regime:
\begin{equation}
\left(\frac{1}{\tau_{\rm{A}}(\varepsilon_{\mathbf{k}})}\right)_{\rm{EP,HT}} =  
\frac{2\beta_A^2 k_BT}{\mu_S \hbar v_{\rm{A}}^2}  
\frac{\varepsilon_{\mathbf{k}}}{(\hbar v_F)^2} \\
\label{eq:tau_A_EP}
\end{equation} 
where $\mu_S=2M/S_{\Re}$ is the mass density per unit area of graphene. The 
full derivation can be found in App. \ref{rhoA}. The effective sound velocity 
for the sum of TA and LA contributions $v_{\rm{A}}$ is such that: 
\begin{equation}
\frac{2}{v_{\rm{A}}^2}=\frac{1}{v_{\rm{TA}}^2}+\frac{1}{v_{\rm{LA}}^2} 
\label{eq:va}
\end{equation}
For the sum of LO and TO phonons, labeled by $\rm{O} \equiv \rm{LO}+\rm{TO}$, 
we use no other approximation than the constant phonon dispersion 
($\hbar \omega_{\rm{LO}}=\hbar \omega_{\rm{TO}}=\hbar \omega_{\rm{O}}=0.20$ eV) 
and find the expression given in Eq. \ref{eq:tauLO/TO} of App. 
\ref{relax-times}.
The same approximation is made for Optical A$_1'$ phonons at K 
($\hbar \omega_{\rm{A}_1'}=0.15$ eV) and we find the expression given in Eq. 
\ref{eq:tauK} of App. \ref{relax-times}. 
Finally, impurity scattering can be easily included knowing the residual 
resistivity $\rho_{\rm{I}}(T=0)=\rho(T=0)$:
\begin{equation}
\frac{1}{\tau_{\rm{I}}(\varepsilon_{\mathbf{k}})} = 
\frac{1}{\tau_{\rm{I}}(\varepsilon_F)}
= \frac{e^2v_F^2 }{2}DOS(\varepsilon_F) \rho_I(T=0) 
\end{equation}

\begin{figure}
\includegraphics{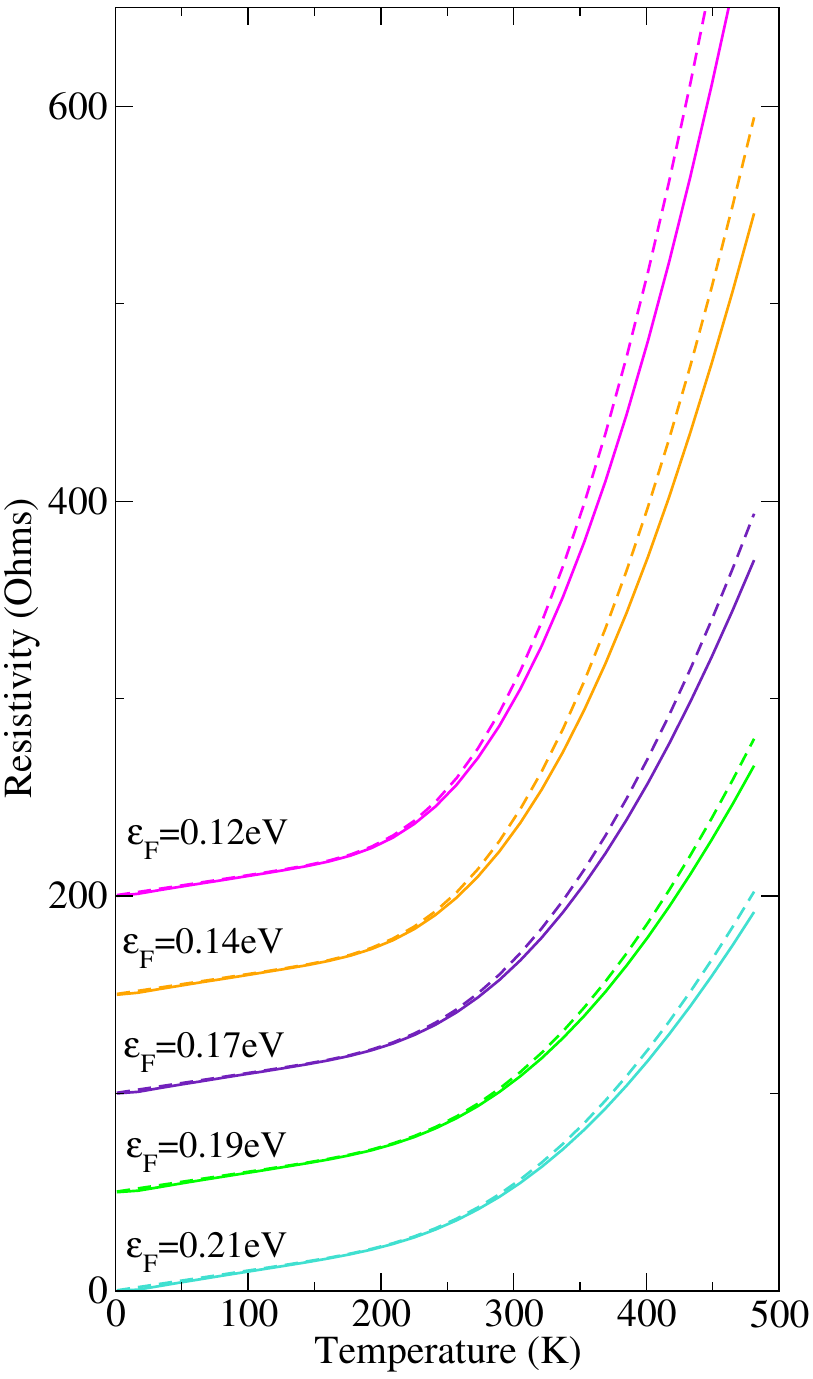}
\caption{(Color online) Comparison between the full Boltzmann transport 
solution (plain lines) and the semi-analytical solution with the 
$\tau(\varepsilon_{\mathbf{k}}) \approx \tau(\varepsilon_{\mathbf{k'}})$ 
approximation made and the expressions of $\tau_{\nu}$ given in Sec. 
\ref{sec:semianal} (dotted lines). Fitted EPC parameters were used. Residual 
resistivity was included in the transport simulation then subtracted for the 
plots. A fictitious residual resistivity was then added for clarity.}
\label{fig:tau_approx}
\end{figure}

Defining $\tau=\left( \sum_{\nu} \frac{1}{\tau_{\nu}}  \right)^{-1}$,  
and numerically evaluating the integral in Eq. \ref{eq:resistivity}
we obtain the results shown in Fig. \ref{fig:tau_approx} that are only
weakly different from the solution of the complete Boltzmann equation. 
The low temperature BG regime ($\rho \propto T^4$) is not reproduced because of 
the quasi-elastic approximation made to obtain Eq. \ref{eq:tau_A_EP}. However, 
a more complicated yet analytical expression for $\tau_{\rm{A}}$ in the BG 
regime is given in App. \ref{relax-times} and yields better results. In the EP 
regime, both solutions are equivalent. The effects of the 
$\tau(\varepsilon_{\mathbf{k'}}) \approx \tau(\varepsilon_{\mathbf{k}})$ 
approximation are seen only slightly in the high temperature regime, when 
optical A$_1'$ phonons dominate the resistivity. It is thus a good and useful 
approximation, since it allows a separate treatment of each contributions and 
the use of Matthiessen's rule. Furthermore, inspecting the times $\tau_{\nu}$ 
validates the statement made in Sec. \ref{sec:EPC-GW}, namely that the 
contribution to resistivity from each phonon is proportional to the squared 
ratio of the EPC parameter and the Fermi velocity.

\subsection{Additivity of resistivities}

\begin{figure}
\includegraphics{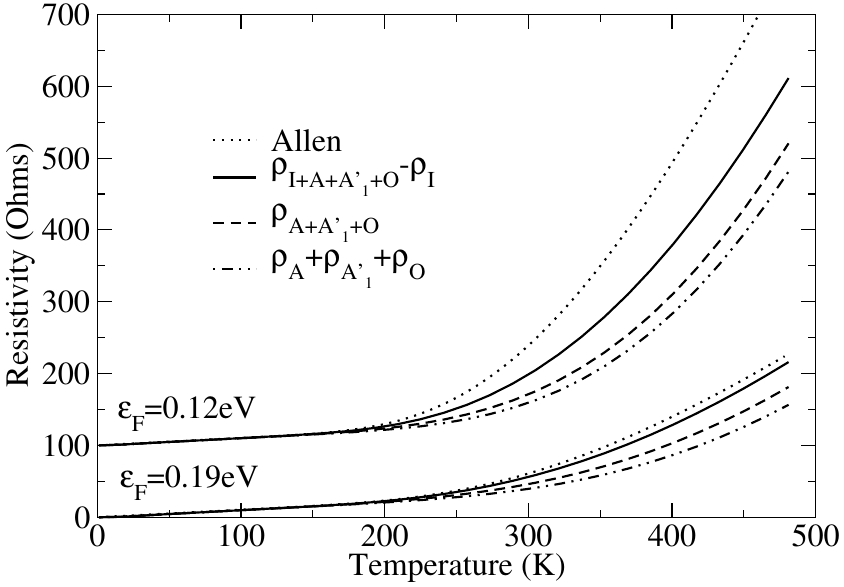}
\caption{Comparison of various methods for the simulation of graphene's 
resistivity. The complete solution to Boltzmann equation (including impurities) 
to which we subtract the residual resistivity 
$\rho_{\rm{I+A+A_1'+O}}-\rho_{\rm{I}}$ gives the result closer to experimental 
conditions. The Boltzmann solution including only phonon contributions 
$\rho_{\rm{A+A_1'+O}}$ corresponds to the theoretical intrinsic resistivity. 
The resistivity made up of the sum of independently derived resistivities is 
$\rho_{\rm{A}}+\rho_{\rm{A_1'}}+\rho_{\rm{O}}$. Allen's method, used on our 
previous work\cite{Park2014}, overestimates the resistivity. Only at low 
temperature are those methods equivalent.}
\label{fig:Additivity}
\end{figure}

As shown in Sec. \ref{sec:semianal}, in the presence of impurities, it
is possible to define independent times $\tau_{\nu}$ for each mode. Then
from each time $\tau_{\nu}$, the
resistivity $\rho_{\nu}$ of a given mode is obtained via the use of
Eq. \ref{eq:resistivity}.
It is then tempting to sum the resistivities to obtain the total
resistivity. However, the energy integral of Eq. \ref{eq:resistivity} should be 
carried on the total time $\tau$, found by adding the {\it inverse} times of 
each modes under Matthiessen's rule. For the resistivities to be additive, it 
is required
that
\begin{equation}
\tau= \left(  \sum_{\nu} \frac{1}{\tau_{\nu}}  \right)^{-1} \approx  \sum_{\nu} 
\left( \frac{1}{\tau_{\nu}}  \right)^{-1}  
\end{equation}
which is rarely valid, as demonstrated in Fig. \ref{fig:Additivity}.
An important consequence is that care needs to be taken when the
resistivity due to impurities (the so called residual resistivity) is
subtracted from the overall resistivity to isolate the intrinsic
contributions. This approach is justified only if the time $\tau_{\rm{I}}$
corresponding to impurities is such that
$1/\tau_{\rm{I}} >> 1/\tau_{\rm{A}}+ 1/\tau_{\rm{A}_1'}+1/\tau_{\rm{O}}$. In 
general, this is not the case and impurity scattering have a more subtle effect 
than just shifting the total resistivity by $\rho_{\rm{I}}$ as shown in Fig. 
\ref{fig:Additivity}. Throughout this work, we include impurity scattering and 
then subtract the residual resistivity. This procedure is convenient but one 
must keep in mind that what remains is not the theoretical intrinsic 
resistivity. Both are plotted in Fig. \ref{fig:Additivity}, as well as Allen's 
method\cite{Allen1978} used in our previous work\cite{Park2014}. The latter 
overestimates the resistivity.

At low temperature, and in the EP regime, the process of adding the residual 
resistivity $\rho_{\rm{I}}(T=0)$ and the acoustic resistivity $\rho_A$ is 
justified and allows one to access directly the magnitude of gauge-field 
parameter $\beta_A$.
Indeed, when only impurities and acoustic phonons
contribute and if $\tau_{\rm{I}}<<\tau_{\rm{A}}$, the corresponding 
resistivities are additive. 
Moreover, in the EP regime, $\rho_{\rm{A}}$ has the simple expression (see 
appendix \ref{rhoA} for detailed derivation):
\begin{equation}
\rho_A \approx \frac{2\pi \beta_A^2 k_BT}{ e^2\hbar v_F^2\mu_S v_{\rm{A}}^2} \
\label{eq:rho_LATA}
\end{equation}
It is clear that the slope of the resistivity
is determined by $v_{\rm{A}}^2$ and $\beta^2_A$. Once the sound velocities are 
known, it is then possible to extract $\beta_A$ directly from transport 
measurements. We expect this parameter to be very close to the amplitude of the 
synthetic vector potential \cite{Low2010}. One must be careful when comparing 
the result of such measurement to other values in literature.
As noted in Ref. \onlinecite{Perebeinos2010}, different EPC models bring 
different pre-factors in Eq. \ref{eq:rho_LATA}. 
For example, the magnitude of the deformation potential $D$ in Ref. 
\onlinecite{Hwang2008a} is defined such that a similar equation is obtained, 
but $D \equiv 2 \beta_A$.

\section{Conclusion}
By diagonalizing the DFT dynamical matrix at finite phonon momentum, we 
developed a model for graphene's acoustic phonon modes.
Based on ab-initio simulations, we demonstrated that inserting those phonon 
modes into the most general symmetry-based model of electron-phonon 
interactions leads to numerically accurate values of acoustic gauge field 
parameter. \\
In order to calculate acoustic EPC parameters in the GW approximation, we 
developed a frozen phonon scheme based on strain deformations. 
We confirmed that the acoustic gauge field is renormalized by electron-electron 
interactions as the Fermi velocity.\\
We then showed that the scattering of electrons by acoustic phonons is mainly 
ruled by the unscreened gauge field while the deformation potential is strongly 
screened. \\
We developed a numerical solution to the complete Boltzmann transport equation 
including  contributions from all phonon branches. Comparison to experiment 
confirm the role of acoustic phonons in the low temperature regime of 
resistivity. In the equipartition regime, the resistivity is proportional to 
$\frac{\beta^2_A}{v^2_A}$ and doping and substrate independent. We found that a 
$15\%$ increase of the acoustic gauge field parameter with respect to the GW 
value gives excellent agreement with experiment. 
In the high temperature regime, scattering by intrinsic A$_1'$ optical phonon 
modes could account for the strong increase in resistivity. 
However, a doping-dependent renormalization of the corresponding EPC parameter 
is necessary, and this renormalization is much stronger than existing estimate 
within GW. The role of remote-phonon scattering at high temperature was not 
ruled out. If remote-phonons are indeed involved, their screening plays an 
important role and needs to be modeled and simulated accurately. \\
We verified the validity of approximations commonly used in the solving of the 
Boltzmann transport equation. An approximate yet accurate semi-numerical 
solution is proposed. Finally, partial analytic solutions were derived in order 
to extract the numerical parameter (gauge field parameter) that relates the 
synthetic vector potential to strain directly from transport measurements. 

\begin{acknowledgments}
The authors acknowledge support from the Graphene Flagship and
from the French state funds managed by
the ANR within the Investissements d'Avenir programme under reference
ANR-11-IDEX-0004-02, ANR-11-BS04-0019 and ANR-13-IS10- 0003-01.
Computer facilities were provided by CINES, CCRT and IDRIS
(project no. x2014091202).
\end{acknowledgments}

\appendix

\section{Numerical Solution to Boltzmann transport equation}
\label{Num}

Eq. \ref{eq:BT1} can be written as a matrix-vector product of the kind 
$\sum_{\varepsilon' } \mathcal{M}_{\varepsilon,\varepsilon'} \tau(\varepsilon') 
= 1 $, 
with $\mathcal{M}  =\mathcal{M}_{\rm{I}} +\sum_{\nu} \mathcal{M}_{\nu} $ where 
$\mathcal{M}_{\rm{I}}$ is just a diagonal matrix containing the inverse 
relaxation times for impurity scattering obtained using existing 
methods\cite{Hwang2007,Hwang2008b,Barreiro2009} and
\begin{eqnarray}
\left[ \mathcal{M}_{\nu} \right]_{\varepsilon,\varepsilon' }&=&  \sum_{\theta'} 
\frac{|\varepsilon'| \Delta \theta' }{(2\pi\hbar v_F)^2 } 
P_{\nu}(\varepsilon,\theta, \varepsilon',\theta')  \frac{  1- 
f^{(0)}(\varepsilon')}{ 1- f^{(0)}(\varepsilon)} \nonumber \\
&& \left( \delta_{\varepsilon,\varepsilon'}      -  
\frac{\cos(\theta')}{\cos(\theta)}    \right)
\label{eq:matdef}
\end{eqnarray}
where angular variables have been discretized with step $\Delta \theta'$ to 
perform numerical integrals, $\theta=\theta_{\mathbf{k}}$ and 
$\theta'=\theta_{\mathbf{k'}}$. The scattering probability 
$P_{\nu}(\varepsilon,\theta, \varepsilon',\theta')$ is the equivalent of Eq. 
\ref{eq:Pep}, defined in a way more suitable for the numerical integration:
\begin{eqnarray}
P_{\nu}(\varepsilon,\theta, \varepsilon',\theta') &=& \frac{2\pi}{\hbar} 
S_{\Re}  |g_{\mathbf{k+q},\mathbf{k},\nu}|^2  
\left\{n_{q,\nu}\delta_{\varepsilon',\varepsilon+\hbar\omega_{\mathbf{q},\nu} } 
\right. \nonumber\\
&+&\left.(n_{q,\nu}+1)\delta_{\varepsilon', 
\varepsilon_{\mathbf{k}}-\hbar \omega_{\mathbf{q},\nu}} \right\} 
\end{eqnarray} 
We represent the matrix 
$\left[ \mathcal{M}_{\nu} \right]_{\varepsilon ,\varepsilon' } $ by discretizing
the energy bands with $N_{\mathcal{M}}=8000$ energy points on a scale of 
$E_{\mathcal{M}}=4\hbar \omega_{\rm{A}_1'}$ around $\varepsilon_F$, such that 
$\Delta \varepsilon= \frac{E_{\mathcal{M}}}{N_{\mathcal{M}}}$.
We sum over matrices associated with each phonon branch. The sum of
scattering probabilities $P_{\rm{LO}} + P_{\rm{TO}}$ being isotropic, 
the corresponding matrix is diagonal. On the contrary, $P_{\rm{TA}}$, 
$P_{\rm{LA}}$ and $P_{A'_1}$ have angular dependencies such that the term 
$\frac{\cos(\theta')}{\cos(\theta)}$ does not integrate to zero and give rise 
to off-diagonal terms.
 
The number of off-diagonal terms in the matrix 
$\left[\mathcal{M}_{\nu} \right]_{\varepsilon ,\varepsilon' } $ depends of the
energy conservation in the scattering probability,
Eq. \ref{eq:Pep}. In the small $|\mathbf{q}|$ limit and for optical A$_1'$
having constant long-wavelength phonon-dispersion, each energy
conservation in Eq. \ref{eq:Pep} is satisfied for only one value of
$\varepsilon $ at fixed phonon momentum. Thus, the
matrix $\left[  \mathcal{M}_{\rm{A}_1'} \right]_{\varepsilon ,\varepsilon' } $ 
has only 2 off-diagonal terms for each $\varepsilon$. 
The energy parametrization is such that the energies 
$\varepsilon \pm \hbar \omega_{\rm{A}_1'}$ are on the grid.

In the case of acoustic phonons, the linear phonon dispersion 
$\hbar \omega_{\mathbf{q},\nu}=v_{\nu} |\mathbf{q}|$ implies that the energy
conservation in Eq. \ref{eq:Pep} is satisfied by a larger subset of
$\varepsilon'$ values for each $\varepsilon$. 
$\left[  \mathcal{M}_{\rm{TA},\rm{LA}} \right]_{\varepsilon ,\varepsilon' } $ 
is thus a band matrix. However, the distance from the diagonal is given by the 
magnitude of the phonon frequency, and since 
$\hbar \omega_{\mathbf{q},\rm{TA}/\rm{LA}}\lesssim \hbar \omega_{2k_F,\rm{LA}}= 
2\frac{v_{\rm{LA}}}{v_F}\varepsilon_F<< \hbar \omega_{\rm{A}_1'}$, 
the band is very narrow compared to the width of the full matrix. 
For the acoustic modes only, we made the approximation that all of off-diagonal 
terms can be summed up and concentrated into the diagonal term, which is 
equivalent to neglecting\cite{Hwang2008a} the variation of $\tau(\varepsilon )$ 
on the energy scale $\hbar \omega_{2k_F,\rm{TA}/\rm{LA}}$. This approximation 
is discussed in App. \ref{relax-times} concerning the acoustic phonons in the 
BG regime. It is not equivalent to the elastic
approximation, as in this case, we do not constrain $\varepsilon =\varepsilon'$ 
in the calculation of each term in Eq. \ref{eq:matdef}.  
\\
Matrix inversion of the $8000\times 8000$ $\mathcal{M}$ matrix gives the time 
$\tau(\varepsilon)$.

\section{Relaxation times}
\label{relax-times}
In this appendix, the 
$\tau(\varepsilon_{\mathbf{k'}}) \approx \tau(\varepsilon_{\mathbf{k}})$ 
approximation is made, such that each phonon mode can be treated separately. 
Some phonon-specific approximations can then be made to simplify the 
calculation of each $\tau_{\nu}$. In the following the indices $A$ and $O$  
designate the summed contributions of acoustic (TA/LA) and optic (LO/TO) 
phonons, respectively.

\subsection{Acoustic phonons in the BG regime:}
The variation of $\tau_{\rm{A}}$ on the scale 
$\hbar \omega_{\mathbf{q},\rm{TA}/\rm{LA}}$ is neglected \cite{Hwang2008a}. 
Since $\hbar \omega_{\mathbf{q},\rm{TA}/\rm{LA}} \ll \varepsilon_F$, the 
initial and final states can be considered to be on the same iso-energetic line 
at $\varepsilon=\varepsilon_F$, which simplifies the angular part of the 
calculus. However, the variation of the electronic occupation must be included 
because $\hbar \omega_{\mathbf{q},\rm{TA}/\rm{LA}}$ is of the order of $k_BT$. 
The following expression of $\tau_{\rm{A}}$ is found, with 
$\nu=\rm{TA},\rm{LA}$:
\begin{eqnarray}
&&\left(\frac{1}{\tau_{\rm{A}}(\varepsilon_{\mathbf{k}})}\right)_{BG}  =
\sum_{\mathbf{k'}} \frac{2\pi}{\hbar} \frac{1}{N}  \sum_{\nu} 
|g_{\mathbf{k'},\mathbf{k},\nu}|^2 
\frac{1-f^{(0)}(\mathbf{k'})}{1-f^{(0)}(\mathbf{k})}  \\
&&\times \big\{ 
n_{|\mathbf{k'}-\mathbf{k}|\nu}
\delta(\varepsilon_{\mathbf{k'}}-\varepsilon_{\mathbf{k}}-
\hbar\omega_{|\mathbf{k}-\mathbf{k'}|,\nu})\nonumber \\
&&+(n_{|\mathbf{k}-\mathbf{k'}|\nu}+1)
\delta(\varepsilon_{\mathbf{k'}}-\varepsilon_{\mathbf{k}}+
\hbar\omega_{|\mathbf{k}-\mathbf{k'}|,\nu}) \big\}
(1-\cos(\theta_{\mathbf{k'}}-\theta_{\mathbf{k}})) \nonumber
\end{eqnarray}
$S_{\Re}$ is the area of the real space unit-cell, $\mu_S$ is the mass density 
per unit area of graphene, and $\mathbf{k'}=\mathbf{k}+\mathbf{q}$. 

\subsection{ Acoustic phonons in the EP and HT regimes: }
The quasi-elastic approximation is valid. The phonon occupation can be 
approximated as 
$n_{q,\rm{TA}/\rm{LA}}\approx\hbar \omega_{\mathbf{q},\rm{TA}/\rm{LA}}/(k_BT)$, 
since $k_BT >> \hbar \omega_{\mathbf{q},\rm{TA}/\rm{LA}}$. We use the following 
expression of $\tau_{\nu}(\varepsilon_{\mathbf{k}})$, easily deduced from Eq. 
\ref{eq:1otaunu} in the elastic case:
\begin{equation}
\frac{1}{\tau_{\nu}(\varepsilon_{\mathbf{k}})}    =  \sum_{\mathbf{k'}} 
P_{\mathbf{k}\mathbf{k'},\nu} \Big(  1  -   
\cos(\theta_{\mathbf{k'}}-\theta_{\mathbf{k}}) \Big) \ \ \text{(Elastic)} 
\end{equation}

The cosine in the scattering probabilities 
$P_{\mathbf{k}\mathbf{k'}, \rm{TA}/\rm{LA}}$ times the cosine in the above 
equation integrates to zero, so that 
$\frac{1}{\tau_{\rm{A}}(\varepsilon_{\mathbf{k}})} = 
\sum_{\nu=\rm{TA},\rm{LA}} \sum_{\mathbf{k'}} P_{\mathbf{k}\mathbf{k'},\nu}$. 
We finally obtain:
\begin{equation}
\left(\frac{1}{\tau_{\rm{A}}(\varepsilon_{\mathbf{k}})}\right)_{\rm{EP,HT}} =  
\frac{2\beta_A^2 k_BT}{\mu_S \hbar v_{\rm{A}}^2}  
\frac{\varepsilon_{\mathbf{k}}}{(\hbar v_F)^2} \\
\end{equation} 
Where $v_{\rm{A}}$ is the effective sound velocity defined in Eq. \ref{eq:va}.

\subsection{Optical LO/TO phonons:}
The scattering probability made by the sum of LO and TO branches 
($P_{\mathbf{k}\mathbf{k'}, \rm{LO}}+P_{\mathbf{k}\mathbf{k'}, \rm{TO}}$) is 
isotropic. We can use a simplified form of Eq. \ref{eq:BT2b} in the isotropic 
case:
\begin{equation}
\frac{1}{\tau_{\nu}(\varepsilon_{\mathbf{k}}) }   =  \sum_{\mathbf{k'}} 
P_{\mathbf{k}\mathbf{k'}, \nu}\frac{  1- f^{(0)}(\varepsilon_{\mathbf{k'}})}
{ 1- f^{(0)}(\varepsilon_{\mathbf{k}})}  \ \  \text{(Isotropic)}
\end{equation}

We obtain the following expression, with 
$\hbar \omega_{\rm{LO}}=\hbar \omega_{\rm{TO}}=\hbar \omega_{\rm{O}}=0.20$ eV, 
and $n_{\rm{LO}}=n_{\rm{TO}}=n_{\rm{O}}=n(\hbar \omega_{\rm{O}})$:

\begin{eqnarray}
&&\frac{1}{\tau_{\rm{O}}(\varepsilon_{\mathbf{k}})}= 
\frac{\beta^2_O}{\mu_S\omega_{\rm{O}}} \frac{1}{(\hbar v_F)^2} \big\{ 
n_{\rm{O}} |\varepsilon_{\mathbf{k}}+\hbar \omega_{\rm{O}} | 
\frac{1-f^{(0)}(\varepsilon_{\mathbf{k}} +
\hbar \omega_{\rm{O}})}{1-f^{(0)}(\varepsilon_{\mathbf{k} }  )} \nonumber \\
&& +(n_{\rm{O}}+1)|\varepsilon_{\mathbf{k}}-\hbar \omega_{\rm{O}} | 
\frac{1-f^{(0)}(\varepsilon_{\mathbf{k}}-
\hbar \omega_{\rm{O}})}{1-f^{(0)}(\varepsilon_{\mathbf{k}})} \big\} 
\label{eq:tauLO/TO}
\end{eqnarray} 

\subsection{Optical A$_1'$ phonons:}

As mentioned before, a difficulty encountered with Optical A$_1'$ phonons at K 
is the change in the EPC matrix element for inter-band scattering. Since we 
consider only electron doping, inter-band scattering occurs only in case of 
phonon emission. We find the general expression of $\tau_{\rm{A}_1'}$ to be, 
with $\hbar \omega_{\rm{A}_1'}=0.15$ eV and 
$n_{\rm{A}_1'}=n(\hbar \omega_{\rm{A}_1'})$  :
\begin{eqnarray}
&&\frac{1}{\tau_{\rm{A}_1'}(\varepsilon_{\mathbf{k}})}= 
\frac{\beta^2_K}{\mu_S\omega_{\rm{A}_1'}} \frac{1}{(\hbar v_F)^2} \\
&&\times \ \ \left\{ \frac{3}{2} n_{\rm{A}_1'} |\varepsilon_{\mathbf{k}}+
\hbar \omega_{\rm{A}_1'} | \frac{1-f^{(0)}(\varepsilon_{\mathbf{k}}+
\hbar \omega_{\rm{A}_1'})}{1-f^{(0)}(\varepsilon_{\mathbf{k} }  )} \right. 
\nonumber  \\
&&\left. +(n_{\rm{A}_1'}+1)  \left( |\varepsilon_{\mathbf{k}}-
\hbar \omega_{\rm{A}_1'} | + \frac{1}{2} ( \varepsilon_{\mathbf{k}}-
\hbar \omega_{\rm{A}_1'}) \right) \right. \nonumber \\
&& \ \ \left. \frac{1-f^{(0)}(\varepsilon_{\mathbf{k}}-
\hbar \omega_{\rm{A}_1'})}{1-f^{(0)}(\varepsilon_{\mathbf{k}})} \right\} 
\nonumber
\label{eq:tauK}
\end{eqnarray}

\section{Derivation of acoustic phonon resistivity 
in equipartition regime \label{rhoA}}

In the EP regime, we can consider scattering by acoustic phonons 
($\nu=\rm{TA},\rm{LA}$) to be elastic:
\begin{equation}
\frac{1}{\tau_{\rm{A}}(\varepsilon_{\mathbf{k}})}    = \sum_{\nu} 
\sum_{\mathbf{k'}} P_{\mathbf{k}\mathbf{k'},\nu} 
\Big(  1  -   \cos(\theta_{\mathbf{k'}}-\theta_{\mathbf{k}}) \Big)
\end{equation}
with
\begin{eqnarray}
P_{\mathbf{k}\mathbf{k'},A}&=&\frac{2\pi}{\hbar} \frac{1}{N} \sum_{\nu} 
|g_{\mathbf{k},\mathbf{k'},\nu}|^2 
\big\{n_{|\mathbf{k'}-\mathbf{k}|\nu}\delta(\varepsilon_{\mathbf{k'}}-
\varepsilon_{\mathbf{k}}-\hbar\omega_{|\mathbf{k}-\mathbf{k'}|,\nu}) 
\nonumber \\
&&+(n_{|\mathbf{k}-\mathbf{k'}|\nu}+1)\delta(\varepsilon_{\mathbf{k'}}-
\varepsilon_{\mathbf{k}}+\hbar\omega_{|\mathbf{k}-\mathbf{k'}|,\nu})\big\} 
\end{eqnarray}
By neglecting the phonon frequency in the delta functions we obtain

\begin{eqnarray}
&&P_{\mathbf{k}\mathbf{k'},A}\approx\frac{2\pi}{\hbar} \frac{1}{N}  \sum_{\nu} 
|g_{\mathbf{k},\mathbf{k'},\nu}|^2  
\delta(\varepsilon_{\mathbf{k'}}-\varepsilon_{\mathbf{k}}) 
\big\{2n_{|\mathbf{k}-\mathbf{k'}|\nu}+1 \big\} \nonumber \\
&&\approx \frac{2\pi}{\hbar} \frac{1}{N} \sum_{\nu=\rm{TA},\rm{LA}} 
\frac{\beta_A^2 k_BT}{\mu_SS_{\Re} v_{\nu}^2}  
(1\pm \cos 3(\theta_{\mathbf{k'}}+\theta_{\mathbf{k}}))  
\delta(\varepsilon_{\mathbf{k'}}-\varepsilon_{\mathbf{k}}) \nonumber
\end{eqnarray}
where $|\mathbf{k}-\mathbf{k'}|=|\mathbf{q}|$ and the angular expressions are 
simplified because $\mathbf{k'}$ and $\mathbf{k}$ are on a iso-energetic line. 
The  $\pm$ sign corresponds to $\rm{LA}$ and $\rm{TA}$ respectively. We made 
the approximation $n_{|\mathbf{q}|,\rm{TA}/\rm{LA}} \approx 
\frac{k_BT}{\hbar \omega_{\mathbf{q},TA/LA}} $, 
since $k_BT >> \hbar \omega_{\mathbf{q},TA/LA}$.

Thus we have:
\begin{widetext}
\begin{eqnarray}
\frac{1}{\tau(\varepsilon_{\mathbf{k}})} &\approx&  \sum_{k'} 
\frac{2\pi}{\hbar} \frac{1}{N}  \sum_{\nu=\rm{TA},\rm{LA}} 
\frac{1}{\mu_SS_{\Re} v_{\nu}^2} \beta_A^2 k_BT 
(1\pm \cos 3(\theta_{\mathbf{k'}}+\theta_{\mathbf{k}}))  
\delta(\varepsilon_{\mathbf{k'}}-\varepsilon_{\mathbf{k}})  
(1-\cos(\theta_{\mathbf{k'}}-\theta_{\mathbf{k}})) \\
\frac{1}{\tau(\varepsilon_{\mathbf{k}})} &\approx& \sum_{\nu=\rm{TA},\rm{LA}} 
\frac{\beta_A^2 k_BT}{\mu_S \hbar v_{\nu}^2}  \frac{1}{(\hbar v_F)^2} \int 
\frac{|\varepsilon_{\mathbf{k'}}|d\varepsilon_{\mathbf{k'}} 
d\theta_{\mathbf{k'}}}{(2\pi)}
(1\pm \cos 3(\theta_{\mathbf{k'}}+\theta_{\mathbf{k}}))  
\delta(\varepsilon_{\mathbf{k'}}-\varepsilon_{\mathbf{k}})  
(1-\cos(\theta_{\mathbf{k'}}-\theta_{\mathbf{k}})) \\
\frac{1}{\tau(\varepsilon_{\mathbf{k}})} &\approx&\sum_{\nu=\rm{TA},\rm{LA}} 
\frac{\beta_A^2 k_BT}{\mu_S \hbar v_{\nu}^2}  
\frac{|\varepsilon_{\mathbf{k}}|}{(\hbar v_F)^2} \int 
\frac{d\theta_{\mathbf{k'}}}{(2\pi)}  
(1\pm \cos 3(\theta_{\mathbf{k'}}+\theta_{\mathbf{k}}))  
(1-\cos(\theta_{\mathbf{k'}}-\theta_{\mathbf{k}})) \\
\frac{1}{\tau(\varepsilon_{\mathbf{k}})} &\approx& 
\frac{2\beta_A^2 k_BT}{\mu_S \hbar v_{\rm{A}}^2}  
\frac{|\varepsilon_{\mathbf{k}}|}{(\hbar v_F)^2}
\end{eqnarray}
\end{widetext}

Doing the usual approximation valid at low temperature:
\begin{eqnarray}
\frac{1}{\rho}&=&\frac{e^2v_F^2}{2} \int d\varepsilon DOS(\varepsilon) 
\tau(\varepsilon) \left( - \frac{\partial f^{(0)}}{\partial \varepsilon} 
(\varepsilon) \right) \\
& \approx &  \frac{e^2v_F^2}{2} DOS(\varepsilon_F) \tau(\varepsilon_F) 
\approx\frac{e^2v_F^2 |\varepsilon_F|}{\pi (\hbar v_F)^2 }  \tau(\varepsilon_F)
\end{eqnarray}
we obtain:
\begin{align}
\rho_A&\approx\frac{\pi (\hbar v_F)^2 }{e^2v_F^2 |\varepsilon_F|} 
\frac{1}{\tau(\varepsilon_F)}\\
\rho_A &\approx  \frac{2\pi \beta_A^2 k_BT}{e^2\hbar v_F^2\mu_S v_{\rm{A}}^2} 
\end{align}

\bibliography{EPC-PRB1}

\end{document}